\documentclass[12pt]{article}
\pdfoutput=1

\usepackage[T1]{fontenc} 
\usepackage{lmodern}
\usepackage{setspace}
\usepackage[hmarginratio=1:1,top=32mm,columnsep=20pt]{geometry} 
\usepackage{fancyhdr} 
\usepackage{abstract}

\usepackage{titlesec} 
\titleformat{\section}[block]{\large\bfseries\centering}{\thesection}{1em}{} 
\titleformat{\subsection}[block]{\bfseries}{\thesubsection}{1em}{} 

\usepackage{color}
\definecolor{dark-gray}{gray}{0.20}
\definecolor{gray}{gray}{0.30}
\definecolor{light-gray}{gray}{0.80}
\definecolor{dark-red}{rgb}{0.7,0,0}
\definecolor{dark-green}{rgb}{0.1,0.4,0}
\definecolor{dark-blue}{rgb}{0.3,0.3,0.7}
\definecolor{light-blue}{rgb}{0.8,0.8,1}

\usepackage[nosort]{cite}
\usepackage{hyperref}
\hypersetup{
	colorlinks=true,
	linkcolor=dark-blue,
	citecolor=dark-red,
	urlcolor=dark-green,
	linktoc=page
}

\usepackage{enumerate}
\usepackage{graphicx}
\usepackage[percent]{overpic}

\usepackage{booktabs}
\usepackage{bbm}
\usepackage[margin=15pt,small]{caption}

\usepackage{amsmath,amssymb,slashed}
\usepackage{amsthm}

\numberwithin{equation}{section}
\usepackage{slashed}

\usepackage{subcaption} 

\newcommand{\dd}{\mathrm{d}}
\newcommand{\dV}{\mathrm{dV}}

\newcommand{\e}{\mathrm{e}}
\newcommand{\rmi}{\mathrm{i}}
\newcommand{\rme}{\mathrm{e}}
\newcommand{\w}{\wedge}

\newcommand{\be}{\begin{equation}}
\newcommand{\ee}{\end{equation}}
\newcommand{\bea}{\begin{eqnarray*}}
\newcommand{\eea}{\end{eqnarray*}}

\newcommand{\f}[2]{\frac{#1}{#2}}

\newcommand{\p}[1]{\phantom{#1}}

\renewcommand{\Re}{\text{Re}}
\renewcommand{\Im}{\text{Im}}

\newcommand{\Tr}{\text{Tr}~}

\newcommand{\N}{\mathcal{N}}

\newcommand{\SU}{\mathop{\rm SU}}
\newcommand{\SO}{\mathop{\rm SO}}

\newcommand{\SL}{\mathop{\rm SL}}
\newcommand{\U}{\mathop{\rm {}U}}
\newcommand{\USp}{\mathop{\rm {}USp}}
\newcommand{\OSp}{\mathop{\rm {}OSp}}
\newcommand{\PSU}{\mathop{\rm {}PSU}}

\renewcommand{\sl}{\mathfrak{sl}}

\def\RR{\mathbb{R}}
\def\bfs#1{\mathbf{#1}}
\def\cals#1{\mathcal{#1}}
\def\eql{~=~}

\title{\vspace{-0mm}\fontsize{24pt}{30pt}\selectfont\textbf{Holographic Interfaces in \\$\mathcal{N}=4$ SYM: Janus and J-folds}\vspace{15mm}}

\author{
\large{
\href{mailto:nikolay.bobev@kuleuven.be}{Nikolay Bobev}$^{J}$, \href{mailto:ffg@kuleuven.be}{Fri{\dh}rik Freyr Gautason}$^{J,j}$ \href{mailto:pilch@usc.edu}{Krzysztof Pilch}$^{\tt{J}}$,}\\[1mm] 
\large{
\href{mailto:minwoosuh1@gmail.com}{Minwoo Suh}$^{\tt{j}}$, and \href{mailto:jesse.vanmuiden@kuleuven.be}{Jesse van Muiden}$^{J}$}\\[6mm]
\normalsize $^{J}$Instituut voor Theoretische Fysica, KU Leuven\\
\normalsize Celestijnenlaan 200D, 3001 Leuven, Belgium\\[3mm]
\normalsize $^{j}$University of Iceland, Science Institute\\
\normalsize Dunhaga 3, 107 Reykjav{\'i}k, Iceland\\[3mm]
\normalsize $^{\tt{J}}$Department of Physics and Astronomy, University of Southern California\\\
\normalsize Los Angeles, CA 90089, USA\\[3mm]
\normalsize $^{\tt{j}}$Department of Physics, Kyungpook National University\\
\normalsize Daegu, 41566, Korea
}

\date{}

\begin{document}

{\hypersetup{urlcolor=black}\maketitle}
\thispagestyle{empty}

\vspace{5mm}

\begin{abstract}

\noindent We find the holographic dual to the three classes of superconformal Janus interfaces in $\mathcal{N}=4$ SYM that preserve three-dimensional $\mathcal{N}=4$, $\mathcal{N}=2$, and $\mathcal{N}=1$ supersymmetry. The solutions are constructed in five-dimensional $\SO(6)$ maximal gauged supergravity and are then uplifted to type IIB supergravity. Corresponding to each of the three classes of Janus solutions, there are also AdS$_4\times S^1\times S^5$ J-fold backgrounds. These J-folds have a non-trivial $\SL(2,\mathbb{Z})$ monodromy for the axio-dilaton on the $S^1$ and are dual to three-dimensional superconformal field theories.
\end{abstract}

\newpage

\setcounter{tocdepth}{2}
\tableofcontents

\section{Introduction}
\label{Sec: Introduction}

Studying quantum field theories with broken Poincar\'e invariance due to the presence of defects or boundaries leads to important insights into their dynamics. The gauge/gravity duality can be applied very effectively in this context to study the physics of such systems. Indeed, co-dimension one defects and interfaces have appeared prominently in holography, see for example \cite{DeWolfe:2001pq} and \cite{Bak:2003jk}. In particular, the duality between type IIB supergravity on AdS$_5\times S^5$ and $\mathcal{N}=4$ SYM can be modified to account for the presence of defects,  \cite{DeWolfe:2001pq}, or interfaces, \cite{Bak:2003jk}. The distinction between these two setups is important. A co-dimension one defect in $\mathcal{N}=4$ SYM supports additional three-dimensional degrees of freedom on its worldvolume while the interface is characterized by position-dependent couplings for the operators in the four-dimensional CFT and no additional degrees of freedom. 

Our goal in this paper is to construct supergravity solutions dual to conformal interfaces, known also as Janus configurations, in $\mathcal{N}=4$ SYM with different amounts of supersymmetry. The original supergravity Janus solution in \cite{Bak:2003jk} breaks all supersymmetry and is invariant under the $\SO(3,2)\times\SO(6)$ subalgebra of the isometry algebra of AdS$_5\times S^5$. The QFT dual to this interface was studied in \cite{Clark:2004sb} where it was also proposed how to construct similar Janus interfaces  preserving $\mathcal{N}=1$ supersymmetry. A more systematic approach to studying superconformal Janus interfaces in $\mathcal{N}=4$ SYM was developed in \cite{DHoker:2006qeo}. It was shown that there are three distinct classes of such interfaces which preserve three-dimensional $\mathcal{N}=4$, $\mathcal{N}=2$, or $\mathcal{N}=1$ supersymmetry, see Table~\ref{Jclass}. This analysis prompted the study of more general co-dimension one deformations of $\mathcal{N}=4$ SYM to include also defect degrees of freedom on the interface compatible with $\mathcal{N}=4$ and $\mathcal{N}=2$ three-dimensional supersymmetry, see \cite{Gaiotto:2008sa,Gaiotto:2008sd,Gaiotto:2008ak}  and \cite{Hashimoto:2014vpa,Hashimoto:2014nwa}, respectively.

%
\begin{table}
\renewcommand{\arraystretch}{1.0}
\centering
\begin{tabular}{@{\extracolsep{15 pt}} cccc}
\toprule
\noalign{\smallskip}
$\N$ & Superalgebra & R-symmetry & Commutant \\
\noalign{\smallskip}
\midrule
\noalign{\smallskip}
4 & $\OSp(4|4,{\mathbb R})$ & $\SU(2)\times\SU(2)$ & \\[4 pt]
2 & $\OSp(2|4,{\mathbb R})$ & $\U(1)$ & $\SU(2)$ \\[4 pt]
1 & $\OSp(1|4,{\mathbb R})$ & &$\SU(3)$\\ 
\noalign{\smallskip}
\bottomrule
\end{tabular}
\caption{\label{Jclass}Three-dimensional $\N =1,2,4$ superconformal algebras that are subalgebras of $\PSU(2,2|4)$. These are the possible symmetry algebras of the superconformal interfaces in $\N=4$ SYM studied in \cite{DHoker:2006qeo}. A subalgebra of the commutant algebra realizes the flavor symmetry of the interface.}
\end{table}

Given the existence of these superconformal Janus interfaces it is natural to look for their holographic description. The AdS$_4$ type IIB supergravity background dual to the $\mathcal{N}=1$ interface of $\N=4$ SYM with $\SU(3)$ flavor symmetry was found in \cite{DHoker:2006vfr}. This solution can also be constructed in five-dimensional $\SO(6)$ maximal gauged supergravity \cite{Clark:2005te}. The five- and ten-dimensional solutions are related by an explicit uplift as shown in \cite{Suh:2011xc}. Moreover, it was found in \cite{Bobev:2019jbi} that this type of $\mathcal{N}=1$ Janus interface exists for all four-dimensional $\mathcal{N}=1$ SCFTs with an AdS$_5$ holographic dual in type IIB supergravity. In \cite{DHoker:2007zhm} it was shown how to find the ten-dimensional supergravity dual of the $\mathcal{N}=4$ Janus interface through a detailed analysis of the supersymmetry variations of type IIB supergravity.

The supersymmetric solutions of type IIB supergravity in \cite{DHoker:2006vfr} and \cite{DHoker:2007zhm} were found by exploiting the large global symmetry of the $\mathcal{N}=1$ and $\mathcal{N}=4$ Janus interfaces. This in turn reduces the BPS supergravity equations to nonlinear PDEs in two variables which can be explicitly analyzed. This strategy is difficult to utilize when studying the $\mathcal{N}=1$ Janus configurations with a smaller flavor algebra or the $\mathcal{N}=2$ Janus interfaces, since the BPS equations become nonlinear PDEs in three or more variables. An alternative approach to circumvent this impasse is offered by five-dimensional gauged supergravity. For the holographic description of Janus interfaces in $\mathcal{N}=4$ SYM one has to study a suitable consistent truncation of the five-dimensional $\SO(6)$ maximal gauged supergravity \cite{Gunaydin:1984qu,Pernici:1985ju,Gunaydin:1985cu} by imposing invariance under the global symmetry group preserved by the interface. This results in a five-dimensional theory with several scalar fields in which one has to construct a supersymmetric domain wall solution with a metric containing an AdS$_4$ factor. The BPS equations for this setup then reduce to several coupled nonlinear ODEs which can be solved analytically or numerically. Once the five-dimensional Janus solution has been constructed, the explicit uplift formulae in \cite{Baguet:2015sma} can be used to find the full type IIB background. Indeed, this approach has been proven useful and supersymmetric Janus solutions in various dimensions  with embedding in string or M-theory were constructed in \cite{Clark:2005te,Bobev:2013yra,Karndumri:2016tpf,Karndumri:2016jls,Karndumri:2017bqi,Gutperle:2017nwo,Suh:2018nmp,Kim:2020unz}.

In this paper we show in detail how to implement this approach to construct the five-dimensional gauged supergravity solutions dual to the $\mathcal{N}=4$ and $\mathcal{N}=2$ Janus interfaces. For completeness we also present the $\mathcal{N}=1$ Janus solution of \cite{Clark:2005te}. We also show explicitly how to uplift all of these solutions to type IIB supergravity. It is worth emphasizing that the $\SU(2)$ and $\SU(3)$ flavor symmetry groups of the $\mathcal{N}=2$ and $\mathcal{N}=1$ interfaces in Table~\ref{Jclass} can be broken while preserving supersymmetry. These less symmetric Janus interfaces were discussed briefly in \cite{DHoker:2006qeo}. Here we will only focus on the more symmetric solutions with $\SU(2)$ and $\SU(3)$ flavor symmetry. We find that our $\mathcal{N}=4$ Janus solution in type IIB supergravity agrees with the solution found in \cite{DHoker:2007zhm}.

\begin{figure}[t]
	\centering
	\begin{subfigure}{.3\textwidth}
		\centering
		\begin{overpic}[scale=0.4]{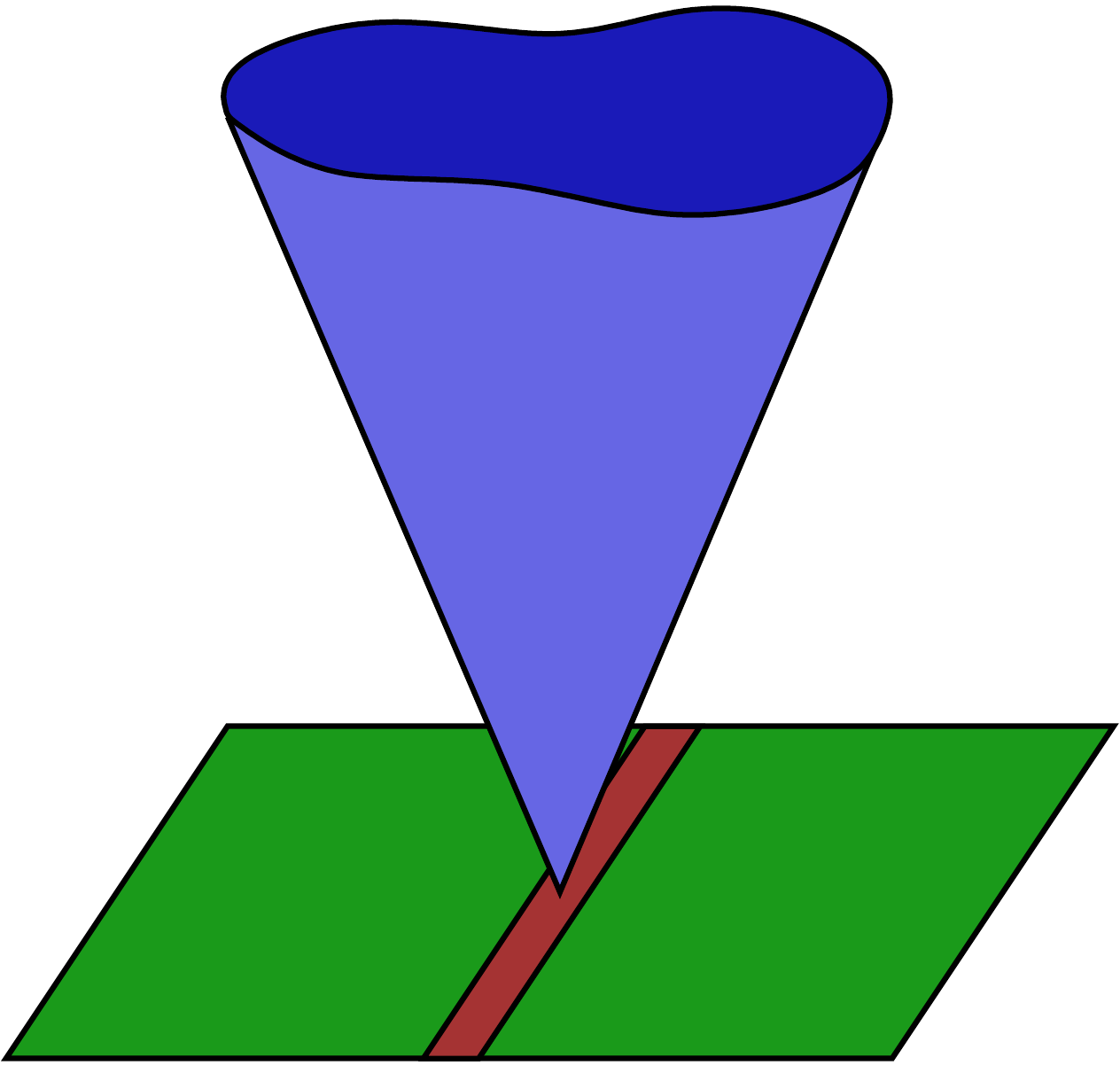}
			\put(43.5,80){\color{white}{\large $\widetilde{S}^5$}}
			\put(28,16.3){{\large$\tau_{\rm L}$}}
			\put(67,16.3){{\large$\tau_{\rm R}$}}
		\end{overpic}
	\end{subfigure}%
	{\hspace{1.8cm}\Huge$ \Rightarrow\quad $}
	\begin{subfigure}{.3\textwidth}
		\centering
		\begin{overpic}[scale=0.35]{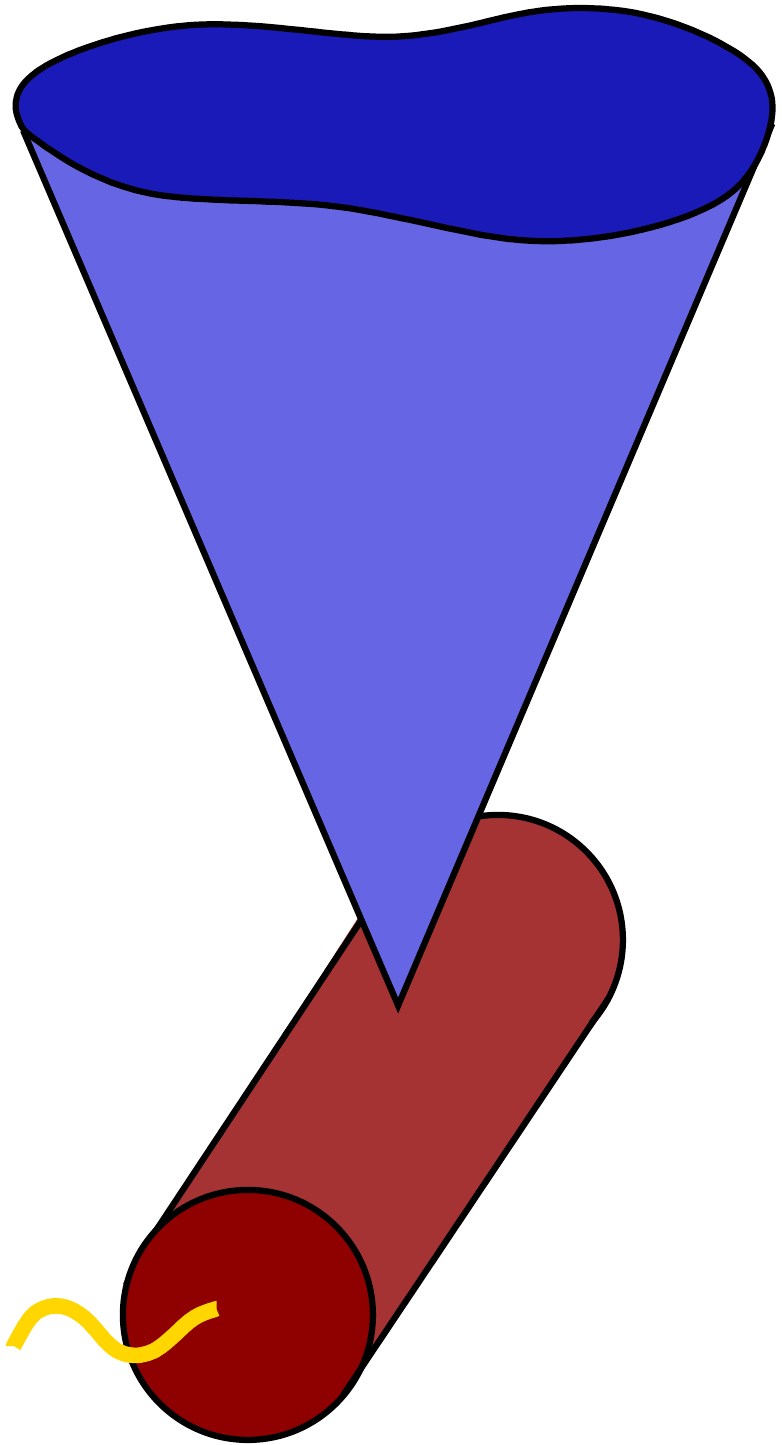}
			\put(22.5,86){\color{white}{\large $\widetilde{S}^5$}}
		\end{overpic}
	\end{subfigure}
	\caption{A schematic illustration of the Janus (left) and J-fold (right) configurations in $\mathcal{N}=4$ SYM realized by D3-branes at the tip of a cone over a deformation of $S^5$.}
	\label{Fig: squashedcone}
\end{figure}

The study of $\tfrac{1}{2}$-BPS Janus interfaces and defects in $\mathcal{N}=4$ SYM led to the discovery of a new class of strongly coupled three-dimensional $\mathcal{N}=4$ SCFTs \cite{Gaiotto:2008sa,Gaiotto:2008sd,Gaiotto:2008ak}. It was shown in \cite{Assel:2018vtq}, see also \cite{Ganor:2014pha,Terashima:2011qi,Gang:2015wya}, how to gauge the global $\U(N)\times \U(N)$ symmetry of these $T[\U(N)]$ theories with a vector multiplet to obtain other strongly coupled three-dimensional SCFTs. While the understanding of these novel three-dimensional QFTs was prompted by the physics of the Janus interfaces, it is important to remember that these are distinct theories and therefore we refer to them as J-fold CFTs. The reason for this moniker becomes more apparent when one studies the holographic description of these J-folds.  As illustrated in Figure~\ref{Fig: squashedcone}, the Janus configurations of $\mathcal{N}=4$ SYM can be realized in type IIB string theory by placing D3-branes at the tip of a cone over $S^5$ and arranging for a non-trivial profile for the axio-dilaton, and the other R-R and NS-NS fields, on the world-volume of the branes. Backreacting this configuration then leads to the AdS$_4$ Janus solutions in supergravity.  Some Janus solutions in supergravity have a linear dilaton profile along the direction transverse to the interface and are therefore strongly coupled asymptotically. It was pointed out in \cite{Inverso:2016eet,Assel:2018vtq} that these backgrounds can be understood as regular S-fold backgrounds of type IIB string theory by compactifying this direction into a circle and imposing a non-trivial $\SL(2,\mathbb{Z})$ monodromy along the $S^1$. It was later shown in \cite{Bobev:2019jbi} how to generalize these $\mathcal{N}=4$ J-folds to AdS$_4$ backgrounds with $\mathcal{N}=1$ supersymmetry. An alternative way to construct J-fold solutions of this type is to employ four-dimensional maximal $[\SO(1,1)\times \SO(6)]\ltimes  \mathbb{R}^{12}$ gauged supergravity. Indeed, the $\mathcal{N}=4$ and $\mathcal{N}=1$ J-fold solutions were found as supersymmetric AdS$_4$ vacua of this theory in \cite{Inverso:2016eet} and \cite{Guarino:2019oct}, respectively.\footnote{We note that four-dimensional gauged supergravity is only suitable for constructing J-fold solutions since regular Janus backgrounds need to be asymptotically AdS$_5$.} As we showed in \cite{Bobev:2019jbi}, yet another way to find J-fold AdS$_4$ solutions is to use five-dimensional gauged supergravity and then uplift them to ten-dimensions and implement the S-fold identification. Here we show that this approach can be systematically implemented with various amounts of supersymmetry and in addition to the $\mathcal{N}=1$ J-fold in \cite{Guarino:2019oct,Bobev:2019jbi}  we also find the $\mathcal{N}=4$ solution of \cite{Inverso:2016eet,Assel:2018vtq} as well as a novel J-fold background with $\mathcal{N}=2$ supersymmetry.

This paper is organized as follows. In Section~\ref{Sec:SO3Janus} we construct the five-dimensional supergravity dual to the $\mathcal{N}=4$ Janus interface and its J-fold and uplift these backgrounds to a solution of type IIB supergravity. We also comment on the relation between our results and those in \cite{DHoker:2007zhm}. In Section~\ref{Sec:SU2U1} the holographic dual to the $\mathcal{N}=2$ Janus interface and its J-fold are presented. We also discuss some properties of the 3d $\mathcal{N}=2$ SCFT dual to the J-fold. The $\mathcal{N}=1$ Janus and J-fold solutions and their type IIB uplifts are briefly presented in Section~\ref{Sec:SU3}. We conclude in Section~\ref{sec:conclusion} with a discussion on some open questions. The appendix contains some details on the gauged supergravity truncations employed in the main text and the derivation of the BPS equations.
\medskip

\noindent
\textit{Note added:} While we were finalizing this paper we became aware of the recent work in \cite{Guarino:2020gfe} which has partial overlap with our results in Section~\ref{subsec:N2Jfold}. The method to obtain the $\mathcal{N}=2$ J-fold solution employed in \cite{Guarino:2020gfe} is based on four-dimensional gauged supergravity and differs from our approach.

\section{The gravity dual of the $\N=4$ interface}
\label{Sec:SO3Janus}

\subsection{The five-dimensional Janus}
\label{ssec:so3janus}

We start by discussing the gravity dual of the $\N=4$ Janus interface. To this end we use the maximal $\SO(6)$ gauged supergravity in five dimensions 
\cite{Gunaydin:1984qu,Pernici:1985ju,Gunaydin:1985cu}. The bosonic sector of the theory consists of the five-dimensional metric, 42 scalar fields, 15 vector fields, and 12 two-forms. This field content is unwieldy to work with and we consider a consistent truncation based on the symmetries preserved by the interface in the field theory. As shown in Table~\ref{Jclass},  the bosonic symmetry algebra is%
\begin{equation}\label{Eq: N4symmetry}
\SO(3,2)\times \SO(3)\times \SO(3)\subset \OSp(4|4,{\mathbb R})~,
\end{equation}
where $\SO(3,2)$ is the conformal algebra preserved by the three-dimensional interface, and $\SO(3)\times\SO(3)$ is the R-symmetry on the interface. Imposing invariance with respect to $\SO(3,2)$ implies that the five-dimensional metric can be written in the form of a curved domain wall
\begin{equation}\label{Eq: 5Dmetric}
\dd s_5^2 = \dd r^2 + \e^{2A(r)} \dd s_{\text{AdS}_4}^2~,
\end{equation}
where $\dd s_{\text{AdS}_4}^2$ is the unit radius metric on AdS$_4$. The vector fields and the 2-forms in the theory have to be set to zero and we are left with a bosonic theory of the metric and scalar fields. 

The scalar fields of the SO(6) gauged supergravity transform in the ${\bf 42}$ of $\USp(8)$ which under  $\SO(6)\times \U(1)_S$ branches to
\be\label{branch42}
{\bf 42}\quad \longrightarrow\quad   {\bf 20}'_0\oplus {\bf 10}_{-2}\oplus \overline{\bf 10}_2\oplus {\bf 1}_4\oplus{\bf 1}_{-4}~.
\ee
These 42 scalar fields have a well-known interpretation in the dual $\mathcal{N}=4$ SYM. The scalars in the ${\bf 20}'$ are dual to the protected scalar bilinear operators in the gauge theory, i.e. all scalar bilinears except the Konishi operator.  The ${\bf 10}$ and $\overline{\bf 10}$ are dual to fermion bilinears and the singlets are dual to the Yang-Mills  coupling and theta angle. Note that $\U(1)_S$ is the compact subgroup of $\SL(2,{\mathbb R})_S$, which is a symmetry of the full scalar potential in five dimensions and is related to the $\SL(2,{\mathbb R})$ symmetry of type IIB supergravity. Its field theory interpretation is the large $N$ avatar of the $\SL(2,{\mathbb Z})$ duality group that acts on the $\N=4$ conformal manifold \cite{Intriligator:1998ig}. 

\begin{table}[t]
\centering
\begin{tabular}{@{\extracolsep{25 pt}}cc}
\toprule
\noalign{\smallskip}
Field & $\SO(6)\times\U(1)_S$ rep \\
\noalign{\smallskip}
\midrule
$\varphi$, $c$ & ${\bf 1}_4\oplus{\bf 1}_{-4}$ \\[4 pt]
$\alpha$ & ${\bf 20}'_{0}$ \\[4 pt]
$\chi$, $\omega$ & ${\bf 10}_{-2}\oplus \overline{\bf 10}_2$ \\ 
\noalign{\smallskip}
\bottomrule
\end{tabular}
\caption{\label{SO3table}The scalar truncation of the maximal supergravity in five dimensions relevant for the holographic dual to the ${\cal N}=4$ interfaces. }
\end{table}

The R-symmetry group of the interface is the block-diagonal  $\rm SO(3)\times SO(3)$ subgroup of $\SO(6)$ under which $\bfs 6\to ({\bf 3},{\bf 1})\oplus ({\bf 1},{\bf 3})$. Using \eqref{branch42}, we find that the truncation with respect to that symmetry results in a simple scalar sector with 
 five fields only that are listed  Table~\ref{SO3table}. In particular, $(\chi,\omega)$ and $(\varphi,c)$ are ${\rm U(1)}_S$ doublets parametrized by the ``moduli'' $\chi$ and $\varphi$, and the phases $\omega$ and $c$, respectively.
The scalar coset in this truncation is 
\begin{equation}\label{Eq: N=4 Scalar manifold}
\f{\SL(3,{\mathbb R})}{\SO(3)}\subset {\rm E_{6(6)}\over \rm USp(8)}\,.
\end{equation}
As we show in Appendix~\ref{appendixSO3}, one can choose the $\rm USp(8)$ gauge such that the scalar   27-bein, $U\in {\rm E}_{6(6)}$, takes the form
\begin{equation}\label{Umso3}
U(\alpha,\chi,\varphi,c,\omega)= V(\alpha,\chi)\cdot U_{\rm SL(2)}(\varphi,c,\omega)\,,
\end{equation}
where $U_{\rm SL(2)}\in \SL(2,\RR)_S$.

The Lagrangian can be written as\footnote{Throughout the paper we work in mostly plus signature  and the action is rescaled with respect to the one in \cite{Gunaydin:1985cu}. In particular, at the $\cals N=8$ supersymmetric vacuum, $\cals P=-3g^2$.}
\begin{equation}\label{Eq:5Dlagrangian}
\mathcal{L} = \frac{1}{16\pi G_N}\sqrt{|g_{5}|}\left(R_{5} +\f1{24}\Tr\left[\partial_\mu M\cdot \partial^\mu M^{-1}\right] - \mathcal{P}\right)~,
\end{equation}
where $M=U^TU$ is the   $\rm USp(8)$ invariant scalar matrix and ${\cal P}$ is the scalar potential.
Our parametrization of the scalar coset in \eqref{Umso3} gives rise to rather complicated kinetic terms in \eqref{Eq:5Dlagrangian}, see \eqref{N4kinterms} in Appendix~\ref{appendixSO3}. However,  due to the $\SL(2,\RR)_S$ invariance, the truncated potential  is very simple and depends only on two of the five scalars. It can be written in terms of a superpotential, $W$, as 
\begin{equation}\label{SO3SO3pot}
\mathcal P = \frac{1}{12}\left| {\partial_\chi W} \right|^2  + \frac{1}{12}\left| {\partial_\alpha W} \right|^2 - \frac{4}{3}\left|  W\right|^2~,
\end{equation}
where 
\begin{equation}
W = -\f{3g}2 \Big(\cosh2\alpha\,\cosh2\chi - \rmi \sinh2\alpha\,\sinh2\chi\Big)~.
\end{equation}

We are interested in $\cals N=4$ supersymmetric Janus solutions of the SO(6) gauged  supergravity, with the metric given by the Ansatz \eqref{Eq: 5Dmetric} and the scalar fields depending on the radial coordinate, $r$, only. For $r\to\pm\infty$, those Janus solutions asymptote to  maximally supersymmetric AdS$_5$ solutions  where  $\alpha=\chi=0$ and the $\SL(2)$-fields $\varphi$, $c$ and $\omega$ are constant but different on  both sides of the interface. Furthermore, from the presence of nontrivial bosonic and fermionic bilinear operators in the dual field theory \cite{DHoker:2006qeo}, we
deduce that the corresponding scalars, $\alpha$ and $\chi$, should both have non-trivial profiles and vanish in the AdS$_5$ asymptotic regions.

As usual, to obtain the required BPS equations we set the fermion supersymmetry variations  \eqref{N8susyvars} to zero. The derivation is quite standard and we have summarized  it  in Appendix~\ref{appendixSO3}. In particular,    the vanishing of the spin-$\tfrac{1}{2}$ variations leads to the following equations\footnote{A  prime denotes the derivative with respect to $r$.}
\begin{equation}\label{Eq: N=4 BPS}
\begin{split}
(\alpha' -\sec(c+2\omega)\varphi')^2 =&\, \frac{1}{36}|\partial_\alpha { W} |^2~,\\[6 pt]
(\chi')(\alpha' -\sec(c+2\omega)\varphi') =&\, \frac{1}{24}\sinh4\chi\,\Re({ W} \partial_\alpha { W} )~,\\[6 pt]
(\varphi')(\alpha' -\sec(c+2\omega)\varphi') =&\, \frac{1}{24}\cos(c+2\omega)\tanh4\chi\, \Im({ W} \partial_\alpha { W} )~,
\end{split}
\end{equation}
and
\begin{equation}\label{sl2eqsso3}
\begin{split}
\omega' =&\, \sinh^2\varphi\,c'~,\\[6pt]
\sinh2\varphi\,c' =&\, -2\tan(c+2\omega)\varphi'~.
\end{split}
\end{equation}
From the spin-$\tfrac{3}{2}$ variations, we obtain two equations for the metric function $A(r)$:
\begin{align}\label{Eq: N=4 metricfunc}
(A')^2 & =\, \frac{1}{9}\left|  { W}  \right|^2-\rme^{-2 A}\,,\qquad 
 \e^{-A}  =\, -\f{\,\Im({ W}\,\partial_\alpha \overline{ W})}{18(\alpha'-\sec(c+2\omega)\varphi')}~.
\end{align}
One can verify that these two equations  are consistent upon using all the BPS equations above, and that solutions to the BPS equations \eqref{Eq: N=4 BPS}-\eqref{Eq: N=4 metricfunc} automatically satisfy the equations of motion  that follow from the Lagrangian \eqref{Eq:5Dlagrangian}.

The construction of the Janus solutions of interest is facilitated by two integrals of motion:
\begin{equation}\label{Eq: N=4 Integrals 1}
{\cal I} \equiv  8\,\f{(\cosh4\alpha\,\cosh4\chi-1)^3}{\sinh^{4}4\chi}~,
\end{equation}
and
\begin{equation}\label{Eq: N=4 Integrals 2}
{\cal J} \equiv \sinh2\varphi\,\sin(c +2 \omega)~,
\end{equation}
which follow  from  \eqref{Eq: N=4 BPS}  and \eqref{sl2eqsso3}, respectively.
For example, using  \eqref{Eq: N=4 BPS}, \eqref{Eq: N=4 Integrals 1} and  
the second equation in \eqref{Eq: N=4 metricfunc}, we obtain an algebraic relation, %
\begin{equation}\label{Eq: N=4 A alg}
\e^{-6A}=\f{g^6}{64\,{\cal I}}\,\sinh^24\chi~.
\end{equation}
Together with \eqref{Eq: N=4 Integrals 1}, this allows us to solve for $\alpha$ and $\chi$ in terms of the metric function,  $A$:\footnote{Throughout the paper, the BPS equations have a discrete symmetry under an overall change of sign  of  some fields. For example,
\eqref{Eq: N=4 BPS}-\eqref{Eq: N=4 metricfunc} are invariant under $\alpha\to-\alpha$, $\chi\to-\chi$, and $\varphi\to-\varphi$. This symmetry leads to a freedom in choosing some signs when solving for some of the fields. To simplify the presentation, we will consistently work with just one set of signs.}
\begin{equation}\label{solalch}
\begin{split}
\cosh 4\alpha & = \frac{2X^2+\mathcal{I}}{2\sqrt{X^4+\mathcal{I}X}}\,,\\[6 pt]
\sinh 4\chi & = \cals I^{1/2} X^{-3/2}\,,
\end{split}
\end{equation}
where we defined
\begin{equation}\label{defX}
X \equiv \frac{g^2 }{4}\,\rme^{2A} \,.
\end{equation}
It follows from the first equation in \eqref{Eq: N=4 metricfunc} and \eqref{Eq: N=4 BPS} that $X$ satisfies a differential equation 
describing the motion of a one-dimensional particle with zero energy in an effective potential:
\begin{equation}\label{Eq: N=4 classmech}
\f{4}{g^2}(X')^2 + V_\text{eff} = 0~,\qquad V_\text{eff} = 4X(1-X)-\mathcal I~.
\end{equation}

The asymptotic AdS$_5$ vacuum solution of the five-dimensional theory is at
\begin{equation}
X\rightarrow +\infty~,\quad \text{such that}\quad \left(\chi,\alpha\right)\rightarrow 0,\quad A \rightarrow +\infty.
\end{equation}
and the Janus solutions we are after interpolate between two such AdS$_5$ regions. From the form of the potential in \eqref{Eq: N=4 classmech} it is clear that  one can find such regular solutions only when $0<\mathcal I\leq 1$. The maximum of the effective potential is at $X=1/2$,
\begin{equation}
V_{\text{eff}}\left(\tfrac{1}{2}\right) = 1-\mathcal I.
\end{equation}
 We thus see that the constant value $X=1/2$ is an exact ``static'' solution whenever $\mathcal I =1$. This will be discussed in more detail in Section~\ref{SO3Jfold} where we focus on the so called J-fold solutions. For a Janus solution, the classical particle comes in from $r=+\infty$ and scatters off the potential back to infinity. The turning point, $r_\text{tp}$, for this scattering is determined  by the largest zero of $V_\text{eff}=0$ located at 
\begin{equation}
X_{\text{tp}} = \frac12 \left(1 + \sqrt{1-\mathcal I}\right)~.
\end{equation}
Integrating \eqref{Eq: N=4 classmech} we obtain
\begin{equation}\label{SO3metricsol}
 X = {1\over 2} \left(1+\sqrt{1-{\cal I}}\cosh(g r-gr_\text{tp})\right)~,
\end{equation}
By shifting the radial coordinate we can set $r_\text{tp}=0$.

\begin{figure}[t]
\centering
\begin{subfigure}{.5\textwidth}
  \centering
  \includegraphics[width=0.95\textwidth]{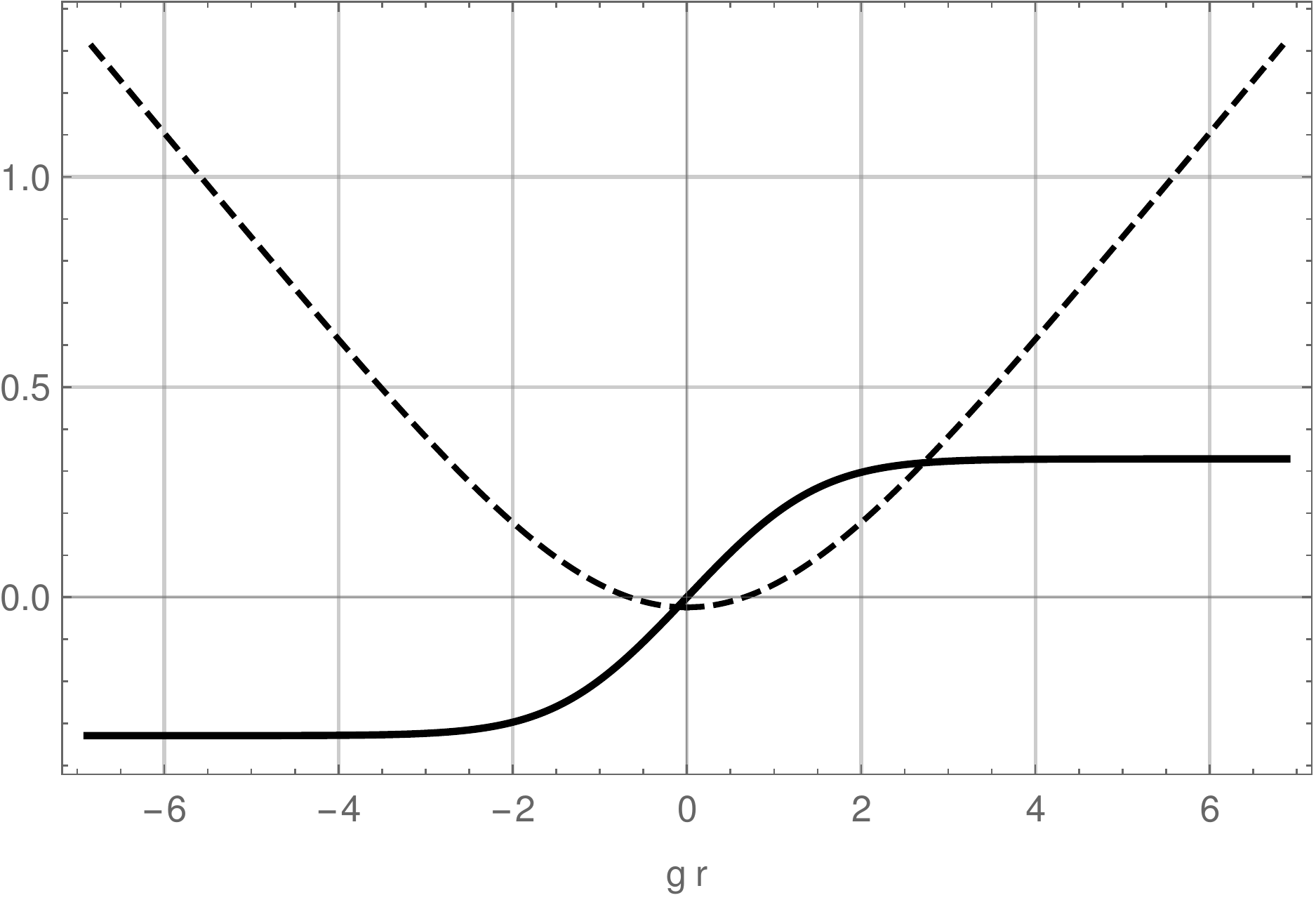}
\end{subfigure}%
\begin{subfigure}{.5\textwidth}
  \centering
  \includegraphics[width=0.95\textwidth]{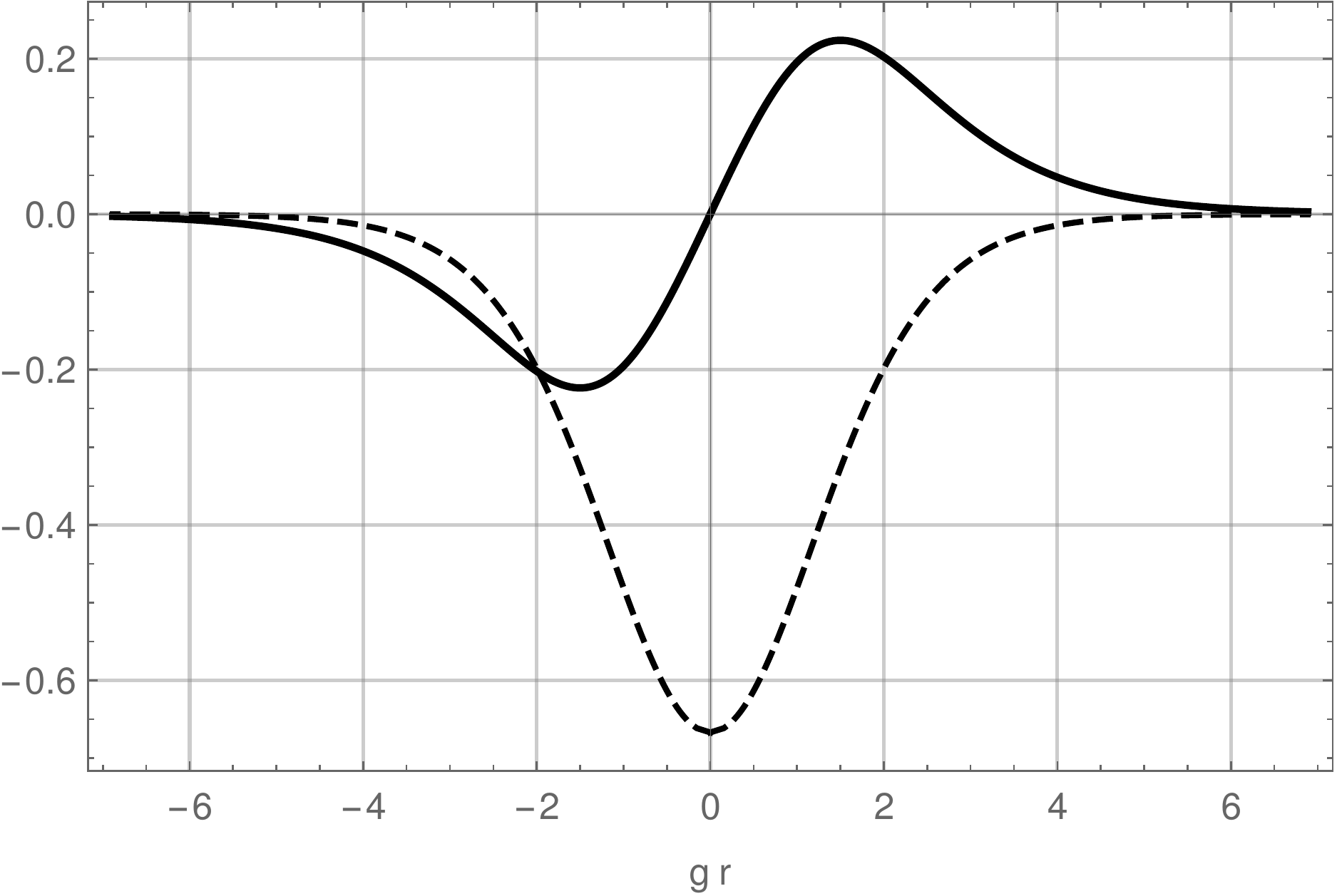}
\end{subfigure}
\caption{\label{SO3SO3plot}A sample solution with ${\cal I}=1/3$. On the left pane we plot the functions $F(r)-F_0$ (solid curve) that controls the solution for $\varphi$ and  $(\log X)/4$ (dashed curve) which through \eqref{defX} represents the warp factor $A$. On the right pane we plot $\sinh4\alpha$ (solid curve) and $\sinh4\chi$ (dashed curve).}
\end{figure}

At this point we  still have to solve for the five-dimensional dilaton, $\varphi$, and the phases $c$ and $\omega$.
By rewriting the third equation in \eqref{Eq: N=4 BPS} in terms of $\varphi(X)$, we find the following ODE
\begin{equation}\label{eq:phiXdiffeq}
\frac{\sinh 2\varphi}{\sqrt{\sinh^2 2\varphi - \mathcal J^2}} \frac{\dd \varphi}{\dd X} =\pm \frac{3 \sqrt{\mathcal I} \left( \mathcal I + 2X^2 \right)}{8 X \left( \mathcal I + X^3 \right)}\frac{1}{\sqrt{-V_{\text{eff}}}}\,,
\end{equation}
which is  integrable and solved by
\begin{equation}\label{Eq: N=4 dilatonsol}
\cosh2\varphi =\cosh2F+\f12\e^{-2F}{\cal J}^2~,
\end{equation}
where 
\begin{equation}\label{Eq: N=4 F function}
F = F_0 \pm \int\limits_{X_{\text{tp}}}^X \f{3\sqrt{{\cal I}}({\cal I}+2x^2)}{8 x({\cal I}+x^3)}\f{\dd x}{\sqrt{-V_\text{eff}(x)}}~,
\end{equation}
and $F_0$ is an integration constant. The integral in \eqref{Eq: N=4 F function} can be evaluated analytically, but the expression is unwieldy and we omit it here. The explicit $r$ dependence is then  obtained using \eqref{SO3metricsol}. However, when the system is studied numerically, it  is more convenient to first use \eqref{eq:phiXdiffeq} to change the integration variable  in \eqref{Eq: N=4 F function} from $X$   to $r$ and then compute  $F$ as a function of $r$ directly.   It is important to note that the sign choice in \eqref{Eq: N=4 F function} is correlated with the one in \eqref{eq:phiXdiffeq} and corresponds to the branch of the square root in \eqref{Eq: N=4 classmech}.  In particular,   in order to obtain a regular solution, one must switch to the opposite sign when the particle in our classical mechanics model passes through the turning point. Indeed, this is one of the characteristic features of any Janus solution \cite{Bak:2003jk,Clark:2004sb,Clark:2005te}. We display a plot of a sample solution in Figure~\ref{SO3SO3plot}.

Equipped with the solution for $\varphi$ and the second integral of motion, $\cals J$, we can integrate the remaining BPS equations \eqref{sl2eqsso3} for $c$ to obtain
\begin{equation}\label{Eq: N=4 csol}
\cos^2(c - c_0) = \f{\sinh^22\varphi-{\cal J}^2}{(1+{\cal J}^2)\sinh^22\varphi}~,
\end{equation}
where $c_0$ is another integration constant. Finally, the solution for the scalar $\omega$ is obtained directly from \eqref{Eq: N=4 Integrals 2} using the explicit solutions for $c$ and $\varphi$ above. 

 We have therefore arrived at a general solution to the system of nonlinear BPS equations \eqref{Eq: N=4 BPS}-\eqref{Eq: N=4 metricfunc} which describes a family of supersymmetric Janus interfaces controlled by five constants
\begin{equation}\label{eq:N4Janusconst}
g~,\qquad F_0~,\qquad{\cal I}~,\qquad {\cal J}~, \qquad c_0~.
\end{equation}
The first of these determines the length scale, $L=2/g$, of   AdS$_5$  and hence the rank  $N$  of the gauge group in the dual $\mathcal{N}=4$ SYM. The constant ${\cal I}$ determines the difference between the value of the dilaton in the two asymptotically AdS$_5$ regions in the solution. In the dual $\mathcal{N}=4$ SYM this translates into the difference in the magnitude of the gauge coupling on the two sides of the interface. The constant $F_0$ in turn controls the sum of these two asymptotic values of the coupling. The constant ${\cal I}$ also determines the magnitude of the leading order terms in the asymptotic expansions of the scalars $\alpha$ and $\chi$, which in turn control  the sources for the dual dimension $2$ and $3$ operators in $\mathcal{N}=4$ SYM. Finally, the constant $c_0$ determines the magnitude of the $\theta$-angle in the dual gauge theory, while ${\cal J}$ controls the asymptotic value of $\omega$, which is dual to the phase of the complex dimension $3$ operator.

\subsection{$\SL(2,{\mathbb R})_S$ transformation of Janus solutions}
\label{Sec:SL2trafo}
As we have emphasized already, the five-dimensional $\SO(6)$ gauged supergravity is invariant under the {\it global\/} $\SL(2,{\mathbb R})_S$ symmetry, which corresponds to  the $\SL(2,{\mathbb R})$ symmetry of type IIB supergravity. In particular, its action on the scalar 27-bein \eqref{Umso3} is given by 
\begin{equation}\label{SL2act}
U \quad \longrightarrow \quad U\cdot \Lambda\,,\qquad \Lambda\in \SL(2,\RR)_S\,.
\end{equation}
The Janus solutions  constructed in Section~\ref{ssec:so3janus} manifestly break that symmetry by the presence of a non-trivial profile for the dilaton $\varphi$, the axion $c$, and $\omega$. We will now argue that the nontrivial action of $\SL(2,\RR)_S$ on those solutions can be used to our advantage to set the integration constants $({\cal J},c_0,F_0)$ to zero. In particular, this implies that without a loss of generality we may set both $c$ and $\omega$ to zero.

From the explicit parametrization of the   scalar 27-bein \eqref{Umso3} given by \eqref{N4cosetparam} in Appendix~\ref{app:N4}, we find that
\begin{equation}\label{sl2grp}
U_{\SL(2)}(\varphi,c,\omega) = \e^{-(c+2\omega)\mathfrak{r}/2}\cdot\e^{\varphi\,\mathfrak{t}}\cdot\e^{(c/2+\pi/4)\mathfrak{r}}\,.
\end{equation}
Consider an $\SL(2,\RR)_S$ transformation, $\Lambda$, such that\footnote{Since  $\SL(2,\RR)_S$ in  \eqref{SL2act} effectively acts only on the $ \SL(2,\RR)$ factor in $U$, we can make the argument quite explicit by working with the corresponding $2\times 2$ matrices,
\begin{equation}\label{}
U_{\SL(2)}(\varphi,c,\omega)= 
\left(\begin{matrix}
 \cos\omega \cosh (\varphi )-\sin (c+\omega ) \sinh \varphi& \cos (c+\omega )
   \sinh (\varphi )-\cosh \varphi\sin\omega \\
 \cosh \varphi\sin\omega+\cos (c+\omega ) \sinh \varphi & \cos\omega
   \cosh  \varphi  +\sin (c+\omega ) \sinh  \varphi 
\end{matrix}\right)\,,
\end{equation}
where we omitted a constant matrix on the right that cancels out from the calculation.
}
\begin{equation}\label{}
U_{\SL(2)}(\varphi,c,\omega)\cdot\Lambda = U_{\SL(2)}(\widetilde \varphi,0,0)\,.
\end{equation}
It is straightforward to check using \eqref{sl2eqsso3} and  \eqref{Eq: N=4 Integrals 2} that $\Lambda$ is indeed constant provided the new dilaton field, $\widetilde \varphi$, satisfies 
\begin{equation}\label{}
{\widetilde \varphi'}={\varphi'\over \sqrt{1-\cals J^2 \mathop{\text{csch}} ^2(2\varphi)}}\,,
\end{equation}
which obviously can be solved.
A more careful analysis using the results in Section~\ref{ssec:so3janus}, in particular  \eqref{Eq: N=4 Integrals 2}, \eqref{Eq: N=4 dilatonsol}, and \eqref{Eq: N=4 csol}, shows that for a given Janus solutions we may simply take
\begin{equation}\label{}
\widetilde \varphi=F\,,
\end{equation}
where the function $F$ is the same as in  \eqref{Eq: N=4 dilatonsol} and $\Lambda$ can be written explicitly  in terms of the constants $\cals J$, $c_0$ and $F_0$.  The solution is then transformed to one with $\cals J=c_0=0$. 
An additional   $\SL(2,\mathbb{R})_S$ transformation can be used to eliminate the integration constant $F_0$, but this does not drastically simplify the Janus configuration any further.

Setting ${\cal J}$ and $c_0$ to zero is a useful simplification and is important for our subsequent discussion. As we have shown above, these parameters can always be reinstated by performing an $\SL(2,\mathbb{R})_S$ transformation  and nothing is lost by setting them to zero. This transformation can also be done at the level of the ten-dimensional solution as we discuss in Section~\ref{subsec:UCLAcomp}.

\subsection{The ten-dimensional Janus}
\label{Sec: The 10D solution SO3 sol}

The five-dimensional Janus solutions above can be uplifted to a solution of ten-dimensional type IIB supergravity \cite{Schwarz:1983qr,Howe:1984sr} using the consistent truncation results in \cite{Baguet:2015sma}.\footnote{See \cite{Bobev:2018hbq,Petrini:2018pjk,Bobev:2018eer,Bobev:2019wnf} for more details and examples on how the uplift formulae of \cite{Baguet:2015sma} are applied to various solutions of the $\SO(6)$ maximal gauged supergravity.} The consistent truncation ensures that supersymmetry is preserved  and that the ten-dimensional equations of motion are satisfied. To present  the ten-dimensional background, 
we choose coordinates on $S^5$ adapted to the $\SO(3)\times \SO(3)$ symmetry of the solution by using the following explicit embedding of $S^5$ in ${\mathbb R}^6$: 
\begin{equation}\label{Eq: Embedding coordinates of five sphere for SO3 sol}
\begin{aligned}
Y^1 =&\cos\theta\cos\phi_1\,,\quad Y^2 =\cos\theta\sin\phi_1\cos\phi_2\,,\quad Y^3 = \cos\theta\sin\phi_1\sin \phi_2\,, \\[6 pt]
Y^4 =&\sin\theta\cos\xi_1\,,\,\,\quad Y^5 = \sin\theta\sin\xi_1 \cos\xi_2\,,\quad\,\, Y^6 =\sin\theta\sin \xi_1\sin\xi_2\,.
\end{aligned}
\end{equation}
Then the  Einstein (round) metric on $S^5$ takes the form
\begin{align}\label{Eq: Round five sphere metric for SO3 sol}
\dd\widehat{\Omega}_{5}^2=\dd\theta^2+\cos^2\theta ~\dd \Omega_2^2+\sin^2\theta ~\dd\widetilde\Omega_2^2~,
\end{align}
where $\dd \Omega_2^2$ and $\dd  \widetilde\Omega_2^2 $ are the metrics on two  $S^2$'s,
\begin{align}\label{eq:2S2}
\dd  \Omega_2^2 =  \dd \phi_1^2 + \sin^2 \phi_1\, \dd \phi_2^2~,\qquad \dd  \widetilde\Omega_2^2 =  \dd \xi_1^2 + \sin^2 \xi_1\, \dd \xi_2^2~,
\end{align}
and the  $\SO(3)\times \SO(3)$ isometry is manifest.
The ten-dimensional solution becomes simpler when written using the following functions:
\be
\begin{split}\label{eq:K1K2N4def}
K_1 =& \sin^2 \theta + \rme^{4\alpha} \cosh 4\chi\,\cos^2 \theta~,\quad \,\, L_1 = \cosh\varphi \, \cos \omega +\sinh \varphi\cos (c + \omega)~,\\[6 pt]
K_2 =& \rme^{-4\alpha} \cosh4\chi\,\sin^2\theta + \cos^2\theta~,\quad L_2 = \cosh \varphi\, \sin \omega + \sinh \varphi \sin( c + \omega )~.
\end{split}
\ee
%
We find that the ten-dimensional metric is\footnote{Throughout this paper we work in Einstein frame. The type IIB conventions are the same as in \cite{Bobev:2018hbq}, or \cite{Polchinski:1998rr} when expressed in string frame.}
\begin{align}\label{Eq: 10D metric N=4}
\dd s^2_{10} = (K_1 K_2)^{1/4} \Bigg( \dd s_5^2 + \frac{4}{g^2} \Big[\dd \theta^2 + \frac{ \cos^2 \theta}{K_1}\dd  \Omega_2^2+ \frac{\sin^2 \theta}{K_2} \dd  \widetilde\Omega_2^2 \Big] \Bigg)~.
\end{align}
The ten-dimensional dilaton and axion are given by 
\begin{equation}\label{Eq: dilaton-axion N=4}
\begin{aligned}
\rme^{-\Phi} &= \frac{\sqrt{K_1 K_2}}{ K_1\, L_1^2\, \rme^{-2\alpha} + K_2\, L_2^2\, \rme^{2\alpha}}~,\\[6 pt]
C_0 &= \frac{K_1 L_1 \left(L_2 -2\cosh \varphi\sin\omega\right) \rme^{-2\alpha}-K_2 L_2 \left(L_1 -2\cosh \varphi\cos\omega\right) \rme^{2\alpha}  }{K_1\, L_1^2\, \rme^{-2\alpha} + K_2\, L_2^2\, \rme^{2\alpha}}~.
\end{aligned}
\end{equation}
%
The two-forms are given by
\begin{equation}\label{eq:B2C2N4Janus}
\begin{aligned}
B_2 + \rmi\, C_2 = \frac{4\sinh 4\chi}{g^2 K_1 K_2}\Big(&\rme^{-\alpha} K_1 \left[ L_1 +\rmi \left(L_2 -2 \cosh \varphi \sin \omega\right) \right]\sin^3 \theta\, \dV_2 \\
&-\, \rme^{\alpha} K_2 \left[L_2 -\rmi\left( L_1 - 2\cosh \varphi \cos\omega \right)\right] \cos^3\theta \, \dV_1\Big)~,
\end{aligned}
\end{equation}
%
where
\begin{equation}\label{eq:V12def}
\begin{aligned}
\dV_1 = \sin \phi_1~\dd\phi_1\w \dd \phi_2~ ,\qquad \dV_2 = \sin \xi_1 ~\dd\xi_1\w\dd \xi_2~,
\end{aligned}
\end{equation}
are the volume forms on the two 2-spheres in \eqref{eq:2S2}. The R-R four-form is given by\footnote{As usual, the $C_4$ form determines only part of the five-form field $F_5$. The full five-form field is then obtained from ${\rm d}C_4$  by imposing the self-duality,  see \cite{Bobev:2018hbq} for further details.}
\begin{equation}\label{eq:C4N4Janus}
\begin{aligned}
C_4 =\frac{2}{g^4} \left(\frac{\sin^3 2\theta}{K_1 K_2} \left( K_1 - K_2 -\tfrac12 \sinh^2 4\chi \cos2\theta \right) + \sin4\theta -4\theta \right)\dV_1 \wedge \dV_2~.
\end{aligned}
\end{equation}
We have fully specified the ten-dimensional Janus solution in terms of the analytic solution for the five-dimensional metric and scalar fields. As a nontrivial  consistency check of the uplift, we have verified that this background  solves the equations of motion of type IIB supergravity.

 \subsection{Comparison to the literature}
 \label{subsec:UCLAcomp}

A ten-dimensional Janus solution with the same isometry and supersymmetry was found in \cite{DHoker:2007zhm}.  Upon comparing the metric and background fluxes in Section~10.3 of \cite{DHoker:2007zhm} with the ones in our uplift of the five-dimensional solution given   in  \eqref{Eq: 10D metric N=4}%
-\eqref{eq:C4N4Janus}, respectively,  we find a complete match between the two solutions. Note that in \cite{DHoker:2007zhm}, the $\SL(2,\mathbb{R})$ symmetry of type IIB supergravity was employed to simplify the ansatz, and, in particular, to set the type IIB axion to zero. This is a ten-dimensional analogue of the five-dimensional argument  in Section~\ref{Sec:SL2trafo}. 

The two solutions can be matched onto each other by the following map between the coordinates $(x,y)$ used in \cite{DHoker:2007zhm} and our coordinates $(r,\theta)$:
 \begin{equation}\label{xyeqs}
\begin{aligned}
 x= \,\frac{g r}{2} + \frac14 \log\left[ \frac{1+\sqrt{\mathcal I}}{1-\sqrt{\mathcal I}} \right]\,, \qquad y=\, \theta\,.
\end{aligned}
 \end{equation}
In addition, the integration constants $(\phi_{+},\phi_{-})$ in \cite{DHoker:2007zhm}, which specify the asymptotic values of the dilaton on the two sides of the interface, and our $(\mathcal{I},F_0)$ are related by:
 \begin{equation}
 \e^{2\phi_{+}-2\phi_{-}} = \frac{1+\sqrt{\mathcal I}}{1-\sqrt{\mathcal I}}\, ,\qquad  \e^{2\phi_{+}+2\phi_{-}}  =\e^{4F_0}\,.
 \end{equation}
Finally, we have to fix the scale of AdS$_4$ used in \cite{DHoker:2007zhm} as $\ell=1$. 

To further facilitate the comparison between the two solutions,  we also note that the functions $K_1$ and $K_2$ defined in \eqref{eq:K1K2N4def} are related to the functions $D$ and $N$ in \cite{DHoker:2007zhm} by
\begin{equation}
\frac{K_2}{K_1} = \frac{\rme^{x} + \rme^{-x}}{\rme^{x} +\e^{2\phi_{+}-2\phi_{-}} \rme^{-x}} \frac{D}{N}\,,
\end{equation}
and two of the five-dimensional supergravity scalars are given in  terms of the coordinates in \cite{DHoker:2007zhm} as 
\begin{equation}\label{varphieqs}
\rme^{2\varphi - 2\alpha} = \e^{2\phi_{+}}\frac{\rme^{x} + \rme^{-x}}{\rme^{x} +\e^{2\phi_{+}-2\phi_{-}} \rme^{-x}} \,.
\end{equation}
The relations in \eqref{xyeqs}-\eqref{varphieqs} fully specify the map between the solution presented in Section~\ref{Sec: The 10D solution SO3 sol} and the one in \cite{DHoker:2007zhm}. 

It is important to realize that the map above is valid for solutions with $\mathcal{J}=c_0=0$ for which the ten-dimensional axion vanishes. As we explained in Section~\ref{Sec:SL2trafo}, the $\SL(2,\mathbb{R})_S$ symmetry of the five dimensional theory can be employed to transform any solution for which $\mathcal{J}$ and $c_0$ are not zero, to one for which both vanish. This transformation can also be done at the level of the ten-dimensional solution, i.e. any solution in Section~\ref{Sec: The 10D solution SO3 sol} with a non-trivial IIB axion, can be transformed to a solution for which the axion vanishes. In fact, the same transformation matrix as used in Section~\ref{Sec:SL2trafo} to set $\mathcal{J}=c_0=0$ can be used in ten dimensions to set the axion to zero.

To study this in more detail it is useful to establish exactly how $\SL(2,\mathbb{R})_S$ relates to the $\SL(2,\mathbb{R})$ symmetry of type IIB. Recall that $\SL(2,\mathbb{R})_S$ acts on the coset element $U$ by a right-multiplication $U\mapsto U\cdot \Lambda$. Therefore the matrix $M$ transforms as
\begin{equation}\label{Mtraforule}
M\mapsto \Lambda^T\cdot M \cdot \Lambda~.
\end{equation}
Next, we note that the explicit relation between the matrix $M$ and the type IIB axion-dilaton matrix $m$ is
\begin{equation}\label{eq:IIBmmatrix}
[m^{-1}]^{\alpha\beta}\equiv m^{\alpha\beta} = \Delta^{4/3}Y_IY_J M^{I\alpha,J\beta}\,.
\end{equation}
Here $Y_I$ are the embedding coordinates in \eqref{Eq: Embedding coordinates of five sphere for SO3 sol} and the function 
\begin{equation}\label{}
\Delta = (K_1K_2)^{-3/8}\,,
\end{equation}
is found by imposing that the determinant of $m^{\alpha\beta}$ equals 1 \cite{Pilch:2000ue}. Also, $\Delta^{-2/3}$ is the warp factor in the metric \eqref{Eq: 10D metric N=4} .  

The index structure of $M$ requires a short explanation. The  matrix $M$ is a $27\times 27$ symmetric matrix which is split into $15\times 15$, $15\times 12$, and $12\times 12$ blocks according to the branching rule ${\bf 27}\to ({\bf 15},{\bf 1}) \oplus ({\bf 12},{\bf 2})$ when E$_{6(6)}$ is broken to $\SL(6,\mathbb{R}) \times \SL(2,\mathbb{R})_S$ (see \cite{Baguet:2015sma} for details). It is the 12$\times$12 block that appears in \eqref{eq:IIBmmatrix} where $I,J=1,\dots,6$ and $\alpha,\beta=1,2$.

To read off the ten-dimensional axion and dilaton from the matrix in \eqref{eq:IIBmmatrix}, we use the standard formula:
\begin{equation}\label{eq:mupdef}
m^{-1}=\left(\begin{array}{cc}
\rme^\Phi & C_0 \rme^{\Phi} \\ 
C_0 \rme^\Phi & \rme^{-\Phi} + C_0^2 \rme^{\Phi}
\end{array} \right)\,.
\end{equation}
Comparing \eqref{eq:IIBmmatrix} and \eqref{Mtraforule} we can translate how $\SL(2,\mathbb{R})_S$ acts on the ten-dimensional fields:
\begin{equation}\label{actonm}
m\mapsto \Lambda^{-1}\cdot m \cdot (\Lambda^{-1})^T~.
\end{equation}
We note that whereas in \eqref{Mtraforule}, $\Lambda$ is an $\SL(2,\mathbb{R})_S$ matrix embedded in E$_{6(6)}$, here it is simply a 2$\times$2 matrix. We now see explicitly how $\SL(2,\mathbb{R})_S$ of the five-dimensional theory relates to the  $\SL(2,\mathbb{R})$ symmetry type IIB supergravity.
Another way to package the action of $\SL(2,\mathbb{R})$ on the type IIB fields, see for example Chapter 12 of \cite{Polchinski:1998rr}, is to define $\tau = C_0+\rmi \e^{-\Phi}$ and write
\begin{equation}\label{SL2Translation}
\left(\begin{array}{c}
C_2  \\  B_2
\end{array} \right)\mapsto 
\left(\begin{array}{cc}
{a} & {b} \\ 
{c}& {d}
\end{array} \right) 
\left(\begin{array}{c}
C_2  \\  B_2
\end{array} \right)\,, \quad \tau' = \dfrac{{a}\,\tau+{b}}{{c}\,\tau+{d}}\,,\quad \text{with}\quad \Lambda = 
\left(\begin{array}{cc}
{d} & {b} \\ 
{c}& {a}
\end{array} \right)\,.
\end{equation}
Using these rules we have explicitly verified that the transformation $\Lambda$ found in Section~\ref{Sec:SL2trafo} can be used at the level of the ten-dimensional solution in Section~\ref{Sec: The 10D solution SO3 sol} to set the axion to zero.

\subsection{An $\N=4$ J-fold}
\label{SO3Jfold}

We now return to five dimensions and study the special solution of the BPS equations with $\cals I=1$, where $\cals I$ is the integral of motion in \eqref{Eq: N=4 Integrals 1}. The effective potential in \eqref{Eq: N=4 classmech} has a critical point at $X=1/2$. For ${\cal I}=1$, the potential energy vanishes at this point, which implies that  $X=1/2$  is a static solution to the classical mechanics problem in \eqref{Eq: N=4 classmech}. This solution is very interesting and we discuss it in some detail below. 


First, using  \eqref{defX} we find that the metric takes the simple form 
\begin{equation}
\dd s_5^2 = \frac{4}{g^2}\left( \dd \rho^2 + \frac12 \dd s_{\text{AdS}_4}^2 \right) 
~,
\end{equation}
where  $\rho = gr/2$. Secondly, \eqref{solalch} implies 
that $\alpha$ and $\chi$ are constant,
\begin{equation}\label{Eq: N=4 Jfoldsol}
\alpha = 0~,\qquad \cosh 4\chi = 3~.
\end{equation}
Finally, in view of the discussion in Section~\ref{Sec:SL2trafo}, we set $c=\omega=0$ upon which the BPS equations 
\eqref{Eq: N=4 BPS}-\eqref{Eq: N=4 metricfunc} collapse to a single equation
\begin{equation}\label{}
\varphi' = {g\over 2}\,,
\end{equation}
whose solution is a linear function,
\begin{equation}\label{}
\varphi=\rho+\varphi_0\,.
\end{equation}


If the coordinate $\rho$ is non-compact, this simple solution is unphysical since the scalar field $\varphi$ blows up as $\rho\to \infty$. As pointed out in \cite{Inverso:2016eet,Assel:2018vtq,Bobev:2019jbi}, one way to remedy this is to compactify $\rho$, such that the  scalar fields become periodic  modulo an   $\SL(2,{\mathbb R})_S$ transformation. To see how this works in detail in our example, recall that $\varphi$ parametrizes the $\SL(2,\RR)_S$ group elements \eqref{sl2grp},
\begin{equation}\label{}
U_{\SL(2)}(\rho)\equiv U_{\SL(2)}(\varphi(\rho)) =  \e^{\rho \mathfrak{t}}\, \e^{(\pi/4)\,\mathfrak{r}}~,
\end{equation}
where we have set $\varphi_0=0$.
Under the translation of coordinate $\rho$ by a period $\rho_0$, we  obviously have
\begin{equation}\label{sl2twist}
U_{\SL(2)}(\rho+\rho_0)= U_{\SL(2)}(\rho)\,\frak J\,,\qquad \frak J\eql \e^{-(\pi/4)\,\frak r}\e^{\rho_0\,\frak t} \e^{(\pi/4)\,\frak r}\eql \e^{-\rho_0\,\frak s}\,.
\end{equation}
Recall that  $\mathfrak{t}$, $\mathfrak{r}$ and $\mathfrak{s}$ are the three generators  of $\SL(2,\RR)_S$ defined in \eqref{trsdef} and hence $\frak J$ is a candidate twist matrix we are looking for. 
The same transformation as in \eqref{sl2twist} holds for the full scalar 27-bein \eqref{Umso3} for this solution. For the scalar matrix $M=U^TU$, we then have
\begin{equation}\label{Mshift}
M(\rho+\rho_0)\eql \frak J^T \,M(\rho)\,\frak J\,.
\end{equation}

The action of $\SL(2,{\mathbb R})_S$ is akin to the $\SL(2,{\mathbb R})$ symmetry of type IIB supergravity. In string theory this symmetry is further broken to $\SL(2,{\mathbb Z})$. Therefore, to ensure that the S-fold identification in \eqref{Mshift} is well defined  we need to quantize the matrix $\mathfrak J$ appropriately. To this end we translate the action on the matrix $M$ in \eqref{Mshift} to ten dimensions using \eqref{eq:IIBmmatrix}, where it simply becomes, see \eqref{actonm}, 
\begin{equation}\label{2dimJ}
m(\rho+\rho_0)\eql \frak J^{-1} \cdot m(\rho)\cdot (\frak J^{-1}) ^T\,,\qquad \frak J\eql \left(\begin{matrix}
\e^{-\rho_0} & 0 \\ 0 & \e^{\rho_0}\,.
\end{matrix}\right)\,.
\end{equation}
Then we must make sure that the twist matrix, $\frak J$, is similar under the global $\SL(2,\RR)$ symmetry to an element in $\SL(2,\mathbb{Z})$. The necessary and sufficient condition for that is that $\frak J$ in \eqref{2dimJ} satisfies
\begin{equation}\label{quantz}
\Tr \frak J\eql 2\cosh\rho_0\equiv n \in \mathbb{Z}\,.
\end{equation}
By an explicit calculation one can check that $\frak J$ is then similar to the canonical matrix
\begin{equation}\label{Jnmatrix}
\frak J_n\eql \left(\begin{matrix}
n & 1\\ -1 & 0
\end{matrix}\right)\,.
\end{equation}
Hence, up to a similarity transformation, the twist matrix $\frak J$ is determined by $\frak J_n$ for some integer $n> 2$. Such a $\frak J_n$ is thus a hyperbolic element of $\SL(2,\mathbb{Z})$.

The procedure outlined above is an alternative way to construct the J-fold solution discussed in \cite{Inverso:2016eet,Assel:2018vtq} as a background in five-dimensional maximal gauged supergravity. To ensure that this solution preserves supersymmetry we have checked explicitly that the five-dimensional supersymmetry parameters are constant as a function of the coordinate $\rho$ and are thus not affected by the periodic identification $\rho\sim\rho+\rho_0$. Moreover, the $\rm USp(8)$ gauge choice for the scalar 27-bein in \eqref{Umso3} is invariant under $\SL(2,\RR)_S$ and hence the twist matrix, $\frak J$, does not act on the fermions. Therefore, our J-fold construction preserves the same number of supersymmetries as the Janus solution in Section~\ref{ssec:so3janus}.

The J-fold solution described above is a good AdS$_4$ vacuum of string theory which   should be dual to a 3d $\mathcal{N}=4$ SCFT. A useful quantity readily computed holographically is the free energy of this SCFT on the round $S^3$. This is captured by the regularized on-shell action of the AdS$_4$ solution computed as in \cite{Emparan:1999pm}, 
\begin{equation}\label{eq:FS3N4onshell}
{\cal F}_{S^3} = \f{\pi L_4^2}{2G_N^{(4)}}\,.
\end{equation}
Here $L_4 =  {\sqrt{2}}/{g}$ is the scale of AdS$_4$ and $G_N^{(4)}$ is the four-dimensional Newton constant, which can be expressed in terms of the Newton constant in five dimensions, $G_N$, 
\begin{equation}\label{Frho0}
\f{1}{G_N^{(4)}} = \f{2\rho_0}{g G_N}\,.
\end{equation}
To express the free energy in \eqref{eq:FS3N4onshell} in terms of microscopic string theory quantities, we also need the relations \cite{D'Hoker:2002aw}
\begin{equation}
 \f{1}{G_N} = \frac{4}{\pi^3g^5\ell_s^8}\,, \qquad N=\frac{4}{\pi g^4\ell_s^4}\,,
\end{equation}
where $\ell_s$ is the string length and $N$ is the number of D3-branes, or alternatively the rank of the gauge group in the dual SCFT. Using these relations as well as \eqref{quantz} in \eqref{eq:FS3N4onshell} and \eqref{Frho0}, we obtain the following free energy
\begin{equation}\label{Eq: Free energy N=4}
{\cal F}_{S^3} = \f{N^2}{2}\text{arccosh}(n/2)\,.
\end{equation}
This result agrees with the expression in \cite{Assel:2018vtq} where the SCFT dual to this J-fold solution was constructed using the $T[\U(N)]$ non-Lagrangian SCFT together with an $\mathcal{N}=4$ $\U(N)$ vector multiplet with Chern-Simons level $n$.

\subsection{The ten-dimensional J-fold}

So far we described the J-fold solution using a five-dimensional perspective. However, we have invoked the ten-dimensional perspective on this solution to constrain the matrix $\mathfrak J$ in \eqref{Mshift} and compute the free energy in \eqref{Eq: Free energy N=4}. Therefore it is also useful to present the full ten-dimensional version of the J-fold background as a solution of type IIB supergravity. The uplift to ten dimensions proceeds as for the Janus interface solution in Section~\ref{Sec: The 10D solution SO3 sol} so we will be brief. 

The ten-dimensional metric is
\begin{equation}
	\dd s^2_{10} = \f{4(w_+ w_-)^{1/4}}{g^2} \left(  \dd \rho^2 + \frac12 \dd s_{\text{AdS}_4}^2 
	+ \dd \theta^2 +\frac{\cos^2 \theta}{w_+}\dd \Omega_2^2 + \frac{\sin^2 \theta}{w_-}\dd \widetilde \Omega_2^2 \right)~,
\end{equation}
where $w_{\pm} = 2 \pm \cos2\theta$. The dilaton and axion are
\begin{equation}
\begin{aligned}
C_0 + \rmi \rme^{-\Phi} = \frac{w_+ \rme^{2\varphi} + w_- \rme^{-2\varphi} +2\rmi \sqrt{w_+ w_-}\sinh \rho_0}{w_+ \rme^{2\varphi -\rho_0} + w_-\rme^{-\left(2\varphi -\rho_0\right)} }~.
\end{aligned}
\end{equation}
The two-forms are
\begin{equation}
\begin{aligned}
B_2 + \rmi C_2 = \frac{8}{g^2\sqrt{\sinh \rho_0}}\Big[ &\xi^+\frac{\cos^3 \theta}{w_+}\dV_1 +\rmi \xi^-\,\frac{\sin^3 \theta}{w_-}\dV_2\Big]~,
\end{aligned}
\end{equation}
where $\xi^\pm = \rme^{\mp\varphi}\left( \rme^{\rho_0/2} \pm \rmi \rme^{-\rho_0/2}\right)$ and the two-forms $\dV_{1,2}$ are defined in \eqref{eq:V12def}. The R-R four-form is given by
\begin{equation}
C_4=\frac{2}{g^4}\left(\frac{3\sin 4\theta}{w_+ w_-}-4\theta\right)\dV_1\wedge\dV_2~.
\end{equation}
We note that this AdS$_4$ solution is subject to the S-fold procedure described above where we take the coordinate $\rho$ to be periodic and act with the $\SL(2,\mathbb{Z})$ matrix $\mathfrak J_n $ as in \eqref{Jnmatrix}.

\section{The gravity dual of the $\N=2$ interface}
\label{Sec:SU2U1}

\subsection{The five-dimensional Janus}
\label{ssec:5dJanus}
We now turn to the supergravity dual of the $\N=2$ interface with $\SU(2)$ flavor symmetry. The construction of this solution proceeds in a similar manner to the one in Section~\ref{ssec:so3janus}.  We start with a consistent truncation of the maximal $\SO(6)$ gauged supergravity by imposing invariance with respect to the bosonic global symmetry of the $\N=2$ Janus interface,
\begin{equation}\label{N2symmetry}
\SO(3,2)\times \U(1)\times \SU(2)\subset \OSp(2|4,{\mathbb R})\times \SU(2)~.
\end{equation}
Here  $\SO(3,2)$ is the conformal group preserved by the three-dimensional interface, the $\U(1)$ is the R-symmetry and the $\SU(2)$ is the flavor symmetry. Invariance with respect to $\SO(3,2)$ implies that the five-dimensional metric can be written as an AdS$_4$ sliced domain wall, see \eqref{Eq: 5Dmetric}.
We can again consistently eliminate the vector and 2-form fields from the five-dimensional supergravity truncation leaving us with a bosonic theory that includes the metric and scalar fields only. The Lagrangian therefore takes the same form as in \eqref{Eq:5Dlagrangian}.
%
%
The $\SU(2)\times\U(1)$ symmetry of the interface also truncates away most of the 42 scalar fields. The embedding of $\SU(2)\times\U(1)$ in $\SO(6)$ goes through the following breaking pattern
\be\label{SU2embedding}
\SO(6) \to \SU(2)_1\times \SU(2)_2 \times \U(1)_{56}\to \SU(2)_1\times \U(1)_R~,
\ee
where $\U(1)_R\subset \SU(2)_2$, and $\SU(2)_1\times \U(1)_R$ is the bosonic symmetry appearing in \eqref{N2symmetry}. This symmetry breaking pattern is fully specified by the branching of the ${\bf 6}$ representation of $\SO(6)$ to $({\bf 2},{\bf 2})_0\oplus ({\bf 1},{\bf 1})_2\oplus ({\bf 1},{\bf 1})_{-2}$, where the subscript denotes the $\U(1)_{56}$ charges, which then branches further to ${\bf 2}_{1}\oplus {\bf 2}_{-1}\oplus {\bf 1}_{0}\oplus {\bf 1}_{0}$, 
%
\begin{table}[t]
\centering
\begin{tabular}{@{\extracolsep{25 pt}}ccc}
\toprule
\noalign{\smallskip}
Field & $\SO(6)\times\U(1)_S$ rep &$\U(1)_{56}$ charge\\
\noalign{\smallskip}
\midrule
\noalign{\smallskip}
$\alpha$ & ${\bf 20}'_{0}$ & 0\\[4 pt]
$\lambda$, $\psi$ & ${\bf 20}'_{0}$ & 4\\[4 pt]
$\chi$, $\omega$ & ${\bf 10}_{-2}\oplus \overline{\bf 10}_2$ & 2\\[4 pt]
$\varphi$, $c$ & ${\bf 1}_4\oplus{\bf 1}_{-4}$ & 0 \\
\noalign{\smallskip}
\bottomrule
\end{tabular}
\caption{\label{SU2table}The scalar truncation of the maximal supergravity in five dimensions relevant for the holographic dual to ${\cal N}=2$ interfaces with $\SU(2)$ flavor symmetry. }
\end{table}
under $\SU(2)_1\times \U(1)_R$. From now on we drop the subscripts on the symmetry groups. 

The scalars in the truncation are listed in Table~\ref{SU2table} and span the scalar manifold
\begin{equation}\label{SU2scalarmfd}
\f{\SO(3,2)}{\SO(3)\times \SO(2)}\times {\mathbb R}_+~,
\end{equation}
where the scalar $\alpha$ lies in ${\mathbb R}_+$.\footnote{Note that even though we use many of the same symbols for the scalar fields in this truncation as in Section~\ref{Sec:SO3Janus}, these scalar fields are not identical inside the maximal gauged supergravity theory and are therefore dual to different operators in $\mathcal{N}=4$ SYM.} The scalar 27-bein, cf.\ \eqref{appUsu2},
\begin{equation}\label{Umsu2}
U(\alpha,\chi,\lambda,\varphi,c,\omega)= V(\alpha,\chi,\lambda)\cdot U_{\rm SL(2)}(\varphi,c,\omega)\,,
\end{equation}
for this coset has a similar structure as in \eqref{Umso3} and is  
discussed further    in Appendix~\ref{appendixsu2}.

 The full scalar kinetic terms for this 7-scalar truncation are given in \eqref{N2kinterms}.
The potential can be written in terms of a superpotential, 
\begin{equation}\label{SU2pot}
\mathcal P = \f14\left| {\partial_\chi W} \right|^2 + \f14\left| {\partial_\lambda W} \right|^2 + \f1{12}\left| {\partial_\alpha W} \right|^2 - \frac{4}{3}\left|  W\right|^2\,,
\end{equation}
where the superpotential is given by
\begin{equation}\label{Eq:SU2U1superpotential}
W = -\frac{g}{2} \e^{-4\alpha}\left( 2 \e^{6\alpha} \cosh 2\chi + \cosh 2\lambda- \rmi \sinh 2\lambda \sinh 2\chi  \right)\,.
\end{equation}

Proceeding as in Section~\ref{ssec:so3janus}, we take the domain wall metric Ansatz in \eqref{Eq: 5Dmetric} and assume that all scalars depend only on the radial coordinate. 
From the vanishing of the spin-1/2 supersymmetry variations, we derive a set of BPS equations,\footnote{See Appendix~\ref{appendixsu2} for more details.} which  naturally split into three groups: the  $(\alpha,\chi,\lambda)$-equations
\begin{equation}\label{eq:abl}
\begin{aligned}
(\alpha')^2 &=\, \f{1}{144}|\partial_\alpha W|^2~,\\
(\chi')(\alpha') &=\,\f1{48}\Re(\partial_\alpha W \partial_\chi\overline W)~,\\
(\lambda')(\alpha') &=\, \f1{48}\left[ \frac{\Im(\partial_\alpha W \partial_\chi\overline W)}{\cosh 2\chi} +  \Re(\partial_\alpha W\partial_\lambda\overline W)\right]~,\\
0&=\, \Im(\partial_\alpha W\partial_\lambda\overline W)~,
\end{aligned}
\end{equation}
the dilaton equation,
\begin{equation}\label{eq:phi}
\begin{aligned}
(\varphi')(\alpha') &=\, \f{\cos(c+2\omega)}{48}\left[\frac{\Im(\partial_\alpha W \partial_\chi\overline W)}{\cosh 2\chi}- \Re(\partial_\alpha W\partial_\lambda\overline W)\right]~,
\end{aligned}
\end{equation}
and the equations for the phases,
\begin{equation}\label{eq:ang}
\begin{aligned}
\omega' &= \,\sinh^2\varphi\,c'~,\\
\sinh2\varphi~c' &=\, -2\tan(c+2\omega)~\varphi'~,\\
\psi'&=\,0~.\\
\end{aligned}
\end{equation}
In addition, vanishing of the spin-3/2 variations yields the following equations:
\begin{equation}\label{metricfuncSU2}
\begin{aligned}
(A' )^2=\, \f19|{\cal W}|^2-\e^{-2A}\,,\qquad \e^{-A}=\, -\f{1}{36(\alpha)'}\Im({\cal W}\,\partial_\alpha \overline{\cal W})\,.
\end{aligned}
\end{equation}

A new feature of these  BPS equations, as compared to  the ones in Section~\ref{ssec:so3janus}, is that the last equation in \eqref{eq:abl},
\begin{equation}\label{Eq: N=2 algebraic 1}
\e^{6\alpha}\cosh2\lambda\,\cosh2\chi=1~,
\end{equation}
 is purely algebraic. It is straightforward to check that \eqref{Eq: N=2 algebraic 1} is consistent with the first three equations in \eqref{eq:abl}, and hence one can use it to eliminate $\lambda$ from \eqref{eq:abl}-\eqref{metricfuncSU2}. The resulting system of equations can now be solved following similar steps as in Section~\ref{ssec:so3janus}. 

As before, there are two integrals of motion, 
\begin{equation}\label{Eq: N=2 algebraic 2}
\mathcal I \equiv -\frac{32 \rme^{6\alpha} \sinh^3 6\alpha}{\sinh^4 2\chi}~,
\end{equation}
and 
\begin{equation}\label{eq:JN2Janus}
{\cal J} \equiv \sinh2\varphi\,\sin(c +2 \omega)~,
\end{equation}
that follow from \eqref{eq:abl} and \eqref{eq:ang}, respectively.
The algebraic equation for the metric in \eqref{metricfuncSU2} is consistent with the other BPS equations as well as the equations of motion and is solved by
\begin{equation}\label{sol2A}
\e^{-2A} =   \f{g^2}{2 \sqrt{\cal I}} \sqrt{-2\e^{2\alpha}\sinh6\alpha}\,.
\end{equation}
Note that \eqref{Eq: N=2 algebraic 1} implies that $\alpha\leq 0$, so that $\cals I>0$ and the metric function is indeed real and positive.

 As in Section~\ref{ssec:so3janus}, the BPS equations can be reduced to a single equation that  describes the dynamics of a one-dimensional particle:
\begin{equation}\label{PartSU2}
\f{4}{g^2}(X')^2 + V_\text{eff} = 0\,,
\end{equation}
where the generalized particle coordinate is\footnote{The choice of the coordinate, $X$, which is different than the one in \eqref{defX}, is dictated by a much simpler form of the resulting effective potential. } 
\begin{equation}\label{}
X = -2 \rme^{6\alpha} \sinh 6\alpha\,,
\end{equation}
and the effective potential is given by
\begin{equation}\label{eq:VeffN2}
\qquad V_{\text{eff}} = - \frac{16\left( 1-X \right)^{1/3} X^2}{\sqrt{\mathcal I}}\left(\sqrt{\mathcal I} - 2 \sqrt{X\left( 1-X \right)} \right)\,.
\end{equation}

The AdS$_5$ vacuum is now found at 
\begin{equation}
X\rightarrow 0~ , \quad \text{such that} \quad \left( \alpha,\chi,\lambda \right) \rightarrow 0~, \quad A\rightarrow \infty~.
\end{equation}
To find regular Janus solutions one must again restrict $0< \mathcal I \leq 1$.  The solutions then start at $X=0$ and bounce off the first zero of the potential wall at the turning point where the potential vanishes,
\begin{equation}
X_{\text{tp}} = \frac{1}{2} \left( 1 - \sqrt{1-\mathcal I}\right)\,,
\end{equation}
and come back to zero. For $\mathcal I=1$, there is again a static solution at $X=1/2$, which is discussed in detail in Section~\ref{subsec:N2Jfold} below. For $0<{\cal I}\leq1$, the classical mechanics problem is solved by 
\begin{equation}\label{classSol}
r(X) = r_\text{tp} \pm \int_{X_\text{tp}}^X \frac{2\dd x}{g\sqrt{-V_\text{eff}(x)}}~,
\end{equation}
where the subscript ``$\text{tp}$'' refers to the turning point. We are free to choose coordinates such that $r_\text{tp}=0$.

The remaining set of BPS equations reduces to the following equation for $\varphi$
\begin{equation}\label{eq:phieqdiffXN2}
\frac{\sinh 2\varphi}{\sqrt{\sinh^2 2\varphi - \mathcal J^2}} \frac{\dd \varphi}{\dd X} =\pm \frac{\left(3 - 2X \right)X}{\sqrt{\mathcal I} \left( 1- X \right)^{4/3} + 2 \left( 1- X \right)^{5/6} \left.X\right.^{3/2}}\frac{1}{\sqrt{-V_{\text{eff}}}}\,.
\end{equation}
While this equation does not seem to admit an analytic solution in terms of known functions, it can be integrated in quadratures, which makes it amenable to numerical analysis. 

\begin{figure}
\centering
\begin{subfigure}{.5\textwidth}
  \centering
  \includegraphics[width=0.95\textwidth]{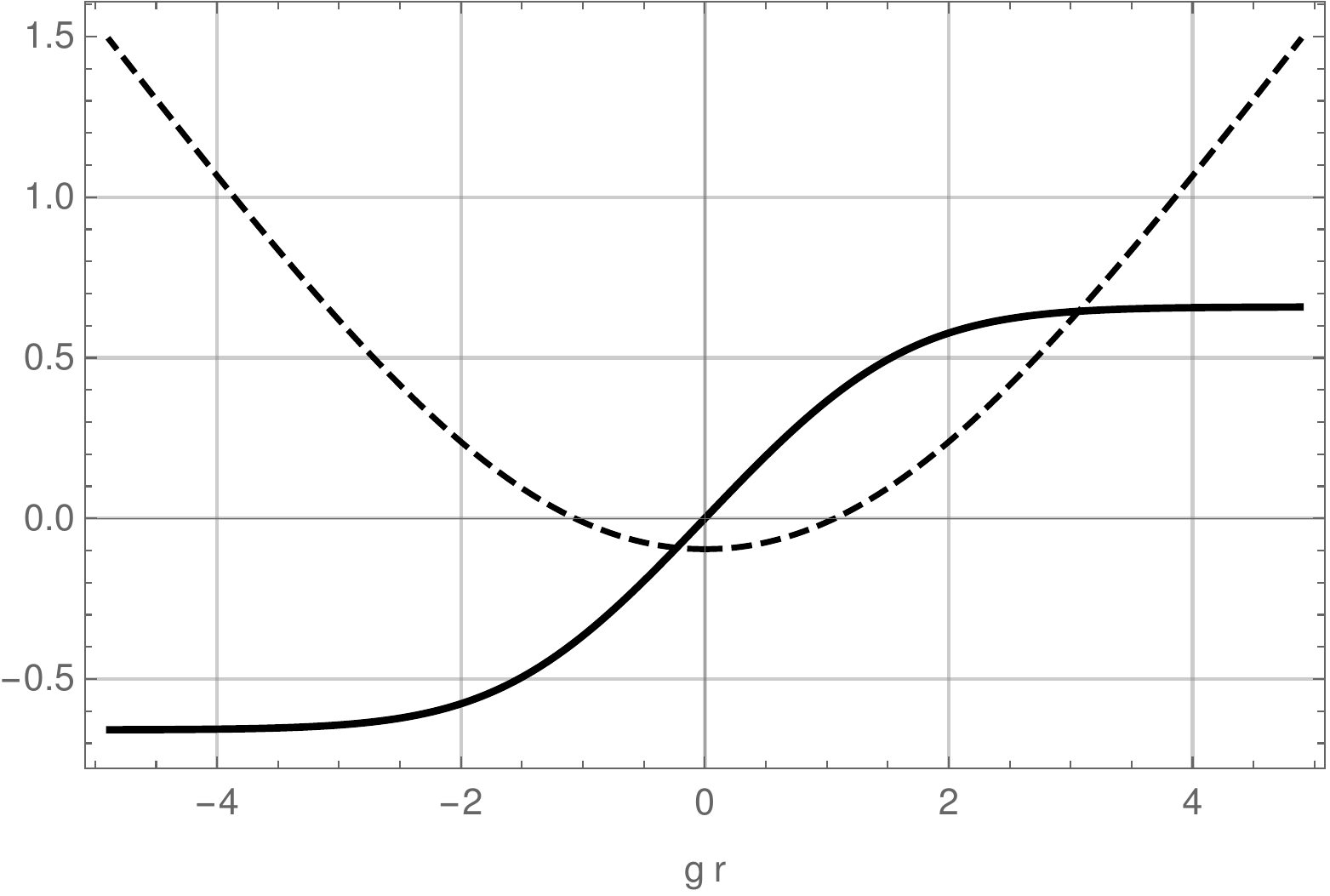}
\end{subfigure}%
\begin{subfigure}{.5\textwidth}
  \centering
  \includegraphics[width=0.95\textwidth]{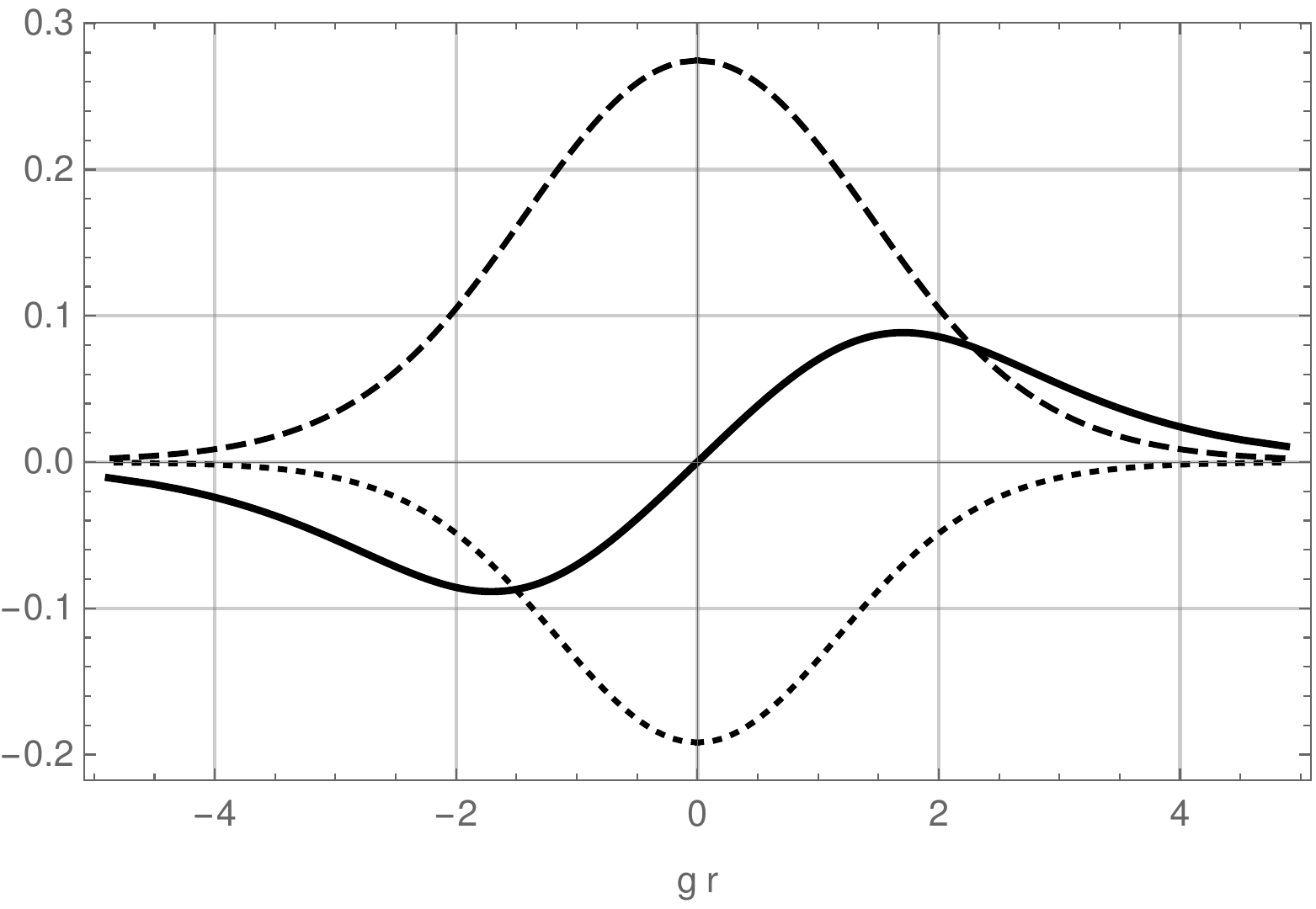}
\end{subfigure}
\caption{\label{SU2U1plot}A sample solution with ${\cal I}=1/3$. On the left pane we plot the functions $F(r)-F_0$ (solid curve) that controls the solution for $\varphi$ and the warp factor $A$ shifted by a constant value (dashed curve). On the right pane we plot $\lambda$ (solid curve), $\chi$ (dashed curve), and $8\alpha$ (dotted curve).}
\end{figure}

The explicit solutions for the five-dimensional dilaton has  the same form as before in \eqref{Eq: N=4 dilatonsol} and \eqref{Eq: N=4 csol}, namely
\begin{equation}\label{eq:coshvarphiN2}
\cosh2\varphi =\cosh2F+\f12\e^{-2F}{\cal J}^2~, 
\end{equation}
where the function $F$ can be determined numerically through the integral
\begin{equation}\label{eq:FfunctN2}
F(X)\equiv F_0 \pm \int\limits_{X_{\text{tp}} }^X \f{(3-2x)x}{\sqrt{\mathcal I}(1-x)^{4/3}+2(1-x)^{5/6}x^{3/2}}\f{\dd x}{\sqrt{-V_\text{eff}(x)}}\,.
\end{equation}
The sign choices in \eqref{eq:phieqdiffXN2} and \eqref{eq:FfunctN2} again reflect the choice of branch when taking the square root in \eqref{PartSU2}. In order to obtain a regular solution we must switch between branches at the turning point of our classical mechanics problem. We display a sample plot of a solution in Figure~\ref{SU2U1plot}. 
The  solution of the axion, $c$, takes the familiar form  
\begin{equation}\label{axionN2}
\cos^2(c - c_0) = \f{\sinh^22\varphi-{\cal J}^2}{(1+{\cal J}^2)\sinh^22\varphi}~.
\end{equation}
Finally, the scalar fields $\chi$ and  $\omega$ are determined using the solutions for $\alpha$, $\varphi$, and $c$ above along with the integrals of motion in \eqref{Eq: N=2 algebraic 2} and \eqref{eq:JN2Janus}, and the angle $\psi=\psi_0$ is constant.

The full $\mathcal{N}=2$ Janus solution is controlled by six constants
\be
g~,\qquad F_0~,\qquad{\cal I}~,\qquad {\cal J}~, \qquad c_0~,\qquad \psi_0~.
\ee
The interpretation of the constants $(g, F_0,{\cal I},{\cal J}, c_0)$ is very similar to that for the $\mathcal{N}=4$ Janus discussed below \eqref{eq:N4Janusconst} with the additional clarification that $\mathcal{I}$ also determines the asymptotic value of the leading term for the scalar $\lambda$ and therefore the magnitude of the source for the complex dimension $2$ operator in the dual $\mathcal{N}=4$ SYM. The phase of this complex operator away from the interface is determined by the asymptotic value for the scalar $\psi$ which is controlled by the integration constant $\psi_0$.

We note that just as for the ${\cal N}=4$ Janus, we can employ the $\SL(2,\mathbb{R})_S$ transformation to greatly simplify the solution and set ${\cal J} = c_0=0$. In fact, the argument is identical to the one presented in Section~\ref{Sec:SL2trafo} and so we will not repeat it here. The broken $\U(1)_{56}$ can be employed to set $\psi_0=0$. These simplifications turn out very useful in the next section where we uplift our solution to ten dimensions.

\subsection{The ten-dimensional Janus}
\label{Sec:SU210D}

Using the uplift formulae in \cite{Baguet:2015sma} we can convert the five-dimensional Janus solutions above to ten-dimensional backgrounds in type IIB supergravity.  While the uplift is essentially algorithmic to execute for general values of the integration constants, in order to keep the expressions for the metric and background fields relatively compact, we will  set
\begin{equation}\label{Eq: N=2 Constantphases}
\mathcal{J}=c_0=0\,,\qquad \psi_0=0~.
\end{equation}
The $\SU(2)\times \U(1)$ symmetry preserved by the five-dimensional solution suggests a convenient coordinate system on $S^5$, which can be written in terms of the embedding coordinates in ${\mathbb R}^6$ as
 \begin{equation}\label{Eq:SU2coords}
\begin{aligned}
 Y^1 +\rmi\, Y^3&=\rme^{-\rmi\, (\xi_1+\xi_2)/2}\cos\theta\cos\xi_3~,\\
 Y^2 + \rmi\, Y^4 &= \rmi\, \rme^{-\rmi\, (\xi_1-\xi_2)/2}\cos\theta\sin\xi_3~,\\
 Y^5+\rmi\, Y^6 &= \rme^{-\rmi\, \phi}\sin\theta~.
\end{aligned}
 \end{equation}
 In these coordinates, the background metric on the round $S^5$ becomes
 \begin{align}\label{Eq: Round five sphere metric}
 \dd\widehat{\Omega}_{5}^2=\dd\theta^2+\sin^2\theta \,\dd\phi^2+\cos^2\theta \left(\sigma_1^2+\sigma_2^2+\sigma_3^2\right)~,
 \end{align}
 where $\sigma_i$ are the $\SU(2)$ left-invariant one-forms obeying the relation $ \dd \sigma_i =  \varepsilon_{ijk} \sigma_j \wedge \sigma_k$ and are given explicitly by
 \begin{align}\label{eq:sig_i}
 \sigma_1 =& - \sin\xi_{1}\dd\xi_3 +   \tfrac{1}{2}\sin 2\xi_3  \cos \xi_{1} \dd\xi_{2} \,,   \notag\\[6 pt]
 \sigma_2 =& - \cos\xi_{1}\dd\xi_3 -  \tfrac{1}{2}\sin  2\xi_3  \sin \xi_{1}\dd\xi_{2}  \,, \\[6 pt]
 \sigma_3 =& -\tfrac{1}{2} \left(\dd\xi_{1}+\cos 2\xi_3 \dd\xi_{2}\right)\,. \notag
 \end{align}
The ranges of the coordinates on $S^5$ are
 \begin{equation}
 \theta\in[0,\pi/2]\,, \qquad \xi_3\in[0,\pi/2]\,, \qquad  \xi_{1},\xi_{2},\phi\in[0,2\pi]\,.
 \end{equation}
 To write down the ten-dimensional solution in a compact form we introduce the following functions:
 \begin{equation}
 \begin{aligned}
 K_1 =& \cos^2 \theta +  C \e^{-6\alpha} b_- \sin^2 \theta\,,\qquad  K_2 = C \cos^2 \theta +  \e^{-6\alpha} b_- \sin^2 \theta\,,\\[6 pt]
 K_3 =& \cos^2 \theta \cosh 2\varphi + C \e^{-6\alpha} d_+ \sin^2 \theta~, \\[6 pt]
 b_{\pm} =& \cosh 2\lambda \pm \sin 2\phi \sinh 2\lambda\,,\\[6 pt]
 d_{\pm} =& \cosh \left(2\lambda \pm 2\varphi\right) - \sin 2\phi \sinh \left(2\lambda \pm 2\varphi\right)\,,\\[6 pt]
 C=& \cosh 2\chi ,\qquad S= \sinh 2\chi~.
 \end{aligned}
 \end{equation}
 The ten-dimensional metric is then given by
 \begin{align}
 \dd s^2_{10} = \e^{\alpha}(C K_1 K_2)^{1/4} \left(\dd s_5^2 + \dd \Omega_5^2 \right)~,
 \end{align}
where $\dd s_5^2$ is the metric in \eqref{Eq: 5Dmetric} and the deformed metric on $S^5$ reads
\begin{equation}\label{Eq:10D metric}
\begin{aligned}
\dd \Omega^2_5 =& \frac{4 \rme^{2\alpha}}{g^2 K_2} \Bigg( \left( b_+  \cos^2\theta + \frac{\e^{-6\alpha} \sin^2 \theta}{C}\right)\dd \theta^2 +\frac{1}{2} \left( b_+ - b_- \right) \frac{ \sin 2\theta}{\tan 2\phi} \,  \dd \theta\, \dd \phi\\
& + b_- \sin^2 \theta\, d\phi^2  + \e^{-6\alpha} \cos^2\theta \left(  \frac{\sigma_1^2}{C} + \frac{K_2}{K_1}\left( \sigma_2^2 + \sigma_3^2 \right)  \right) \Bigg)~ .
\end{aligned}
\end{equation}
The ten-dimensional dilaton and axion are respectively
 \begin{equation}
 \begin{aligned}
 \rme^{\Phi} =& \f{1}{2\sqrt{C K_1 K_2}}\left( 2C K_3 + \e^{-6\alpha} \sin^2 \theta \left(d_- - C^2  d_+ + S^2 \cos2\phi \right) \right)~,\\[6 pt]
 C_0 =& \frac{2 C \cos^2\theta \sinh 2\varphi - \frac12 \e^{-6\alpha} \sin^2 \theta\, \partial_\lambda \left(  d_- - C^2 d_+ \right) }{2C K_3 +  \e^{-6\alpha} \sin^2 \theta \left( d_- - C^2 d_+ + S^2 \cos 2\phi \right)}~.
 \end{aligned}
 \end{equation}
The NS-NS and R-R two-forms can be compactly written as a complex two-form
\begin{equation}
\begin{aligned}
B_2 + \rmi  C_2 =&\frac{2 S \cos \theta}{g^2 K_1K_2}\Big( \frac{2 a_1K_1}{C} \dd \theta \w \sigma_1 \\
& + \sin 2\theta \left( a_+K_1\, \dd \phi \w \sigma_1 -\rmi a_- \e^{-6\alpha} K_2 \, \sigma_2 \w \sigma_3\right)\Big)~,
\end{aligned}
\end{equation}
where
\begin{equation}
\begin{aligned}
a_1=& \rme^{\rmi \phi} \left( \cos^2\theta \sinh\left(\lambda + \varphi\right)C - \sin^2\theta \sinh\left(\lambda - \varphi\right)\e^{-6\alpha} \right)\\
&\qquad + \rmi \rme^{-\rmi\phi} \left( \cos^2\theta \cosh\left(\lambda+\varphi\right) C + \sin^2\theta  \cosh\left(\lambda-\varphi\right)\e^{-6\alpha} \right)\,,\\[6 pt]
a_{\pm} =&  \,\rme^{-\rmi \phi} \cosh\left(\lambda \pm \varphi\right) +\rmi \rme^{\rmi \phi} \sinh\left(\lambda \pm \varphi\right)~.
\end{aligned}
\end{equation}
The R-R four-form is given by
\begin{equation}
\begin{aligned}
C_4 = \frac{8\cos^4\theta}{g^4 K_1 K_2} \Big(\e^{-6\alpha} (K_1 +&C K_2) \sinh 2\lambda \tan\theta\cos 2\phi \,\dd \theta \\
&- (C K_1 + K_2) \dd \phi \Big) \w \sigma_1 \w \sigma_2 \w \sigma_3~ .
\end{aligned}
\end{equation}

Note that although we have set  the scalar $\psi$ to zero for simplicity, one can reintroduce it by performing the following coordinate shift on all ten-dimensional fields above
\begin{equation}
\phi \rightarrow \phi - \frac{\psi}{2}~.
\end{equation}
The reason that in ten dimensions $\psi$ can be reintroduced through such a coordinate transformation is that in the field theory this scalar is related to the phase of a scalar bilinear operator that could be shifted using an $\U(1)_{56}\subset \SO(6)$ rotation. 

The background presented above has an $\SU(2)$ isometry under which the supersymmetry generators are not charged. We can thus construct an orbifold of this supergravity solution by a discrete subgroup of $\SU(2)$ while still preserving the same amount of supersymmetry \cite{Kachru:1998ys}. For a $\mathbb{Z}_k$ orbifold, this construction will preserve a $\U(1)$ subgroup of the $\SU(2)$ flavor symmetry, however, for generic $D$ or $E$ type orbifolds all continuous flavor symmetry will be broken. The holographically dual description of these orbifold solutions should correspond to $\mathcal{N}=2$ superconformal Janus interfaces in the quiver gauge theory obtained by an $ADE$ orbifold of $\mathcal{N}=4$ SYM.

\subsection{An ${\cal N}=2$ J-fold}
\label{subsec:N2Jfold}

Now we go back to five dimensions in order to analyze the solutions with the special value of the integration constant, ${\cal I}=1$. The analysis is analogous to the one in Section~\ref{SO3Jfold} so we will be brief. The effective potential in \eqref{eq:VeffN2} has an extremum at $X=1/2$ for ${\cal I}=1$ where it exactly vanishes. This implies that there is a static solution of the classical mechanics problem for which $X=1/2$. When we express this solution in terms of the five-dimensional supergravity fields, we find that three of the scalars attain constant values
\be
\e^{12\alpha}=\f12~,\qquad \lambda=0~,\qquad \cosh4\chi = 3\,.
\ee
Using the $\SL(2,\mathbb{R})_S$ symmetry of the theory, we can  set $\mathcal{J}=c_0=0$ as explained in Section 2.2. We are also free to set $\psi_0=0$ using the $\U(1)_{56}$ gauge transformation.
The metric is then
\begin{equation}
\dd s_5^2 = \f{2^{4/3}}{g^2} \left( \dd \rho^2 +  \dd s_{\text{AdS}_4}^2 \right)\,,
\end{equation}
where we have introduced a new radial variable $\rho = \frac{g }{2^{2/3}}r$. The dilaton $\varphi$ takes the linear form $\varphi=\rho + \varphi_0$. To obtain a regular solution, we are again forced to periodically identify the radial coordinate $\rho \sim \rho+\rho_0$ and accompany this with an $\SL(2,\mathbb{Z})$ transformation. In order to make this background a good solution of string theory, we have to ensure a proper quantization of the matrix in this S-fold procedure. The result is that we again have to use the same $\SL(2,\mathbb{Z})$ matrix as in \eqref{Jnmatrix}.

We have thus arrived at an AdS$_4$ J-fold solution and following the same steps that lead to \eqref{Eq: Free energy N=4} we can evaluate the  free energy on $S^3$ of the dual $\mathcal{N}=2$ SCFT. The result is
\be\label{eq:FS3N2}
{\cal F}_{S^3} = \f{N^2}{2} \text{arccosh}(n/2)~,
\ee
and we discuss it further in Section~\ref{subsec:N2SCFT}. 

This five-dimensional AdS$_4$ J-fold solution can be uplifted to a background of type IIB supergravity. The metric takes the form
\begin{equation}
\dd s^2_{10} =\frac{2^{\tfrac{3}{2}} w^{\tfrac{1}{4}}}{2^{\tfrac{1}{16}}g^2} \left[ \dd \rho^2 +  \dd s_{\text{AdS}_4}^2  + \dd \theta^2 + \sin^2 \theta \dd \phi^2 + \cos^2 \theta \left( \sigma_1^2 +\frac{2(\sigma_2^2 + \sigma_3^2)}{w}  \right)\right]\,,
\end{equation}
where $w = 1 + \sin^2 \theta$. The dilaton and axion are
\begin{equation}
\begin{aligned}
\rme^\phi =& \frac{\left(w+1\right)\cosh(2\varphi-\rho_0) + \left(w-1\right) \left( \cos2\phi -\sin 2\phi \,\sinh(2\varphi-\rho_0)\right)}{2\sqrt{w}\sinh \rho_0}~,\\
C_0 =& \frac{\left(w+1\right) \cosh2\varphi + \left(w-1\right) \left( \cos 2\phi \cosh \rho_0 - \sin 2\phi \sinh2\varphi \right)}{\left(w+1\right)\cosh(2\varphi-\rho_0) + \left(w-1\right) \left( \cos2\phi -\sin 2\phi \,\sinh(2\varphi-\rho_0)\right)}\,.
\end{aligned}
\end{equation}
The two-form potentials are
\begin{equation}
\begin{aligned}
B_2 + \rmi C_2 &= \frac{2^{\tfrac{7}{16}} \cos\theta}{g^2 \sqrt{\sinh \rho_0}} \Big[ \frac12 \sin2\theta \left( \sin\phi (\xi^+-\rmi \xi^-) + \cos\phi ( \xi^+ + \rmi \xi^- ) \right)\dd \phi \w \sigma_1 \\ 
&+ \left( \sin\phi (\xi^+ + \rmi \xi^-) - \cos\phi ( \xi^+ -\rmi \xi^- )\right) \left( \dd \theta \w \sigma_1 - \frac{\sin 2\theta}{w} \sigma_2 \w \sigma_3 \right) \Big]\,,
\end{aligned}
\end{equation}
where we have defined $\xi^{\pm} = \rme^{\mp\varphi}\left( \rme^{\rho_0/2} \pm \rmi \rme^{-\rho_0/2} \right)$. Finally, the R-R four-form is given by
\begin{equation}
C_4=\frac{4\times2^{\tfrac{7}{8}}\cos^4\theta}{g^4}\left(\frac{1+w}{w}\right) \sigma_1 \w \sigma_2 \w \sigma_3 \w \dd \phi\,.
\end{equation}
%

\subsection{${\cal N}=2$ SCFT intermezzo}
\label{subsec:N2SCFT}

The J-fold AdS$_4$ solution in Section~\ref{subsec:N2Jfold} should be dual to a three-dimensional $\mathcal{N}=2$ SCFT with $\SU(2)$ flavor symmetry. Following the analysis in \cite{Assel:2018vtq} and the result for the holographic free energy in \eqref{eq:FS3N2}, we will now attempt  to identify this SCFT.

It was proposed in  \cite{Assel:2018vtq} that the three-dimensional SCFT dual to the J-fold AdS$_4$ solution presented in Section~\ref{SO3Jfold} can be obtained by taking the strongly coupled $T[\U(N)]$ theory of Gaiotto and Witten \cite{Gaiotto:2008sd} and gauging its global $\U(N)\times \U(N)$ symmetry by an $\U(N)$ $\mathcal{N}=4$ vector multiplet. One also has to add a Chern-Simons term at level $n$ for the gauge field, where $n$ is the integer appearing in the $\SL(2,\mathbb{Z})$ matrix in \eqref{Jnmatrix}. The addition of this Chern-Simons term breaks the manifest supersymmetry in this construction to $\mathcal{N}=3$. However, it was argued in \cite{Assel:2018vtq} that supersymmetry is enhanced to $\mathcal{N}=4$ at the IR fixed point. Further support for this proposal was provided by an explicit calculation of the $S^3$ free energy of this model using supersymmetric localization. The result of this calculation agrees with the holographic free energy in \eqref{Eq: Free energy N=4}. Given these results it is natural to expect that the $\mathcal{N}=2$ SCFT dual to the J-fold solution in Section~\ref{subsec:N2Jfold} can be obtained by a deformation of the construction in \cite{Assel:2018vtq}. In order to identify this theory it is important to note that, despite the significantly different supergravity solutions, the holographic $S^3$ free energy for the $\mathcal{N}=2$ J-fold solution \eqref{eq:FS3N2} is the same as its $\mathcal{N}=4$ counterpart \eqref{Eq: Free energy N=4}.  

One possible explanation for the fact that the free energies of the $\mathcal{N}=4$ and $\mathcal{N}=2$ SCFTs are the same is that the two theories are related by an exactly marginal deformation. Exactly marginal operators are $Q$-exact with respect to the supercharge used for the supersymmetric localization calculation of the $S^3$ free energy, see \cite{Pestun:2016zxk} for a review. This in turn implies that the localization calculation of the $\mathcal{N}=4$ theory performed in \cite{Assel:2018vtq} should also yield the same result for the $\mathcal{N}=2$ J-fold free energy. To establish whether this is the correct procedure to construct the SCFT dual to the $\mathcal{N}=2$ J-fold background  in Section~\ref{subsec:N2Jfold}, one would have to classify the exactly marginal operators in the $\mathcal{N}=4$ SCFT of \cite{Assel:2018vtq}. Since the operator spectrum of this theory is not known, this is currently an open problem. An alternative strategy can be pursued via holography. If there is an exactly marginal deformation that connects the $\mathcal{N}=4$ and $\mathcal{N}=2$ SCFTs, one could attempt to construct  its supergravity dual. This should be realized by a family of AdS$_4$ supersymmetric vacua of IIB supergravity which interpolate between the $\mathcal{N}=4$ and $\mathcal{N}=2$ J-fold solutions in Sections \ref{SO3Jfold}  and \ref{subsec:N2Jfold}. It would be very interesting to either construct these solutions explicitly or rule out their existence.

A three-dimensional SCFT with $\mathcal{N}=2$ supersymmetry and the same free energy as in \eqref{eq:FS3N2} can also be constructed in a different way. One can start with the $T[\U(N)]$ $\mathcal{N}=4$ SCFT and gauge its global $\U(N)\times \U(N)$ symmetry with an $\mathcal{N}=2$ vector multiplet with a Chern-Simons term at level $n$. The free energy of the resulting $\mathcal{N}=2$ IR fixed point can be computed by supersymmetric localization as in \cite{Assel:2018vtq}. Despite the fact that we have modified the theory in \cite{Assel:2018vtq}, the supersymmetric localization calculation will result in exactly the same value for the free energy. To understand this one can decompose the $\N=4$ vector multiplet into an $\N=2$ vector multiplet and an $\N=2$ chiral multiplet in the adjoint representation of the gauge group. As explained in \cite{Kapustin:2009kz,Kapustin:2010xq} one can then show that the contribution of the adjoint chiral multiplet does not affect the supersymmetric localization calculation of the $S^3$ path integral. 

We have therefore arrived at two alternative SCFTs scenarios which explain why the $S^3$ free energies of the $\mathcal{N}=2$ and $\mathcal{N}=4$ Janus solutions in \eqref{eq:FS3N2} and \eqref{Eq: Free energy N=4} are the same. It would be most interesting to understand which of the two proposals outlined above leads to the correct field theory dual of the AdS$_4$ vacuum in Section~\ref{subsec:N2Jfold}.

\section{The gravity dual of the $\N=1$ interface}
\label{Sec:SU3}

In this section we discuss the $\N=1$ Janus and J-fold solutions, with an $\SU(3)$ flavor symmetry, embedded in the maximal gauged $\SO(6)$ supergravity theory, and uplift these solutions to type IIB supergravity. These Janus solutions were previously studied in \cite{Clark:2005te,DHoker:2006vfr,Suh:2011xc}, and we will therefore be brief. In \cite{Bobev:2019jbi}, we have recently constructed a broader class of such $\N=1$ Janus and J-fold solutions embedded in the minimal $\N=2$ gauged supergravity theory coupled to one hypermultiplet, dual to an infinite class of $\N=1$ quiver gauge theories. The corresponding type IIB backgrounds in \cite{Bobev:2019jbi} are of the form AdS$_5 \times M_5$ for the Janus solutions and AdS$_4 \times S^1 \times M_5$ for the J-fold solutions, where  $M_5$ is a  generic Sasaki-Einstein manifold with a squashed metric. By specifying $M_5$ to be $S^5$, one recovers the solutions that are discussed here. The J-fold solution of this type was also recently studied in \cite{Guarino:2019oct} using four-dimensional gauged supergravity, see also \cite{Lust:2009mb} for a local form of this J-fold solution.

\subsection{The five-dimensional Janus}
\label{ssec:5dJanussu3}
Imposing the $\SU(3)$ symmetry truncates the scalar sector of the maximal supergravity theory to only four scalars that parametrize the coset
\begin{equation}
\frac{\SU(2,1)}{\SU(2)\times \U(1)}\,,
\end{equation}
and are listed in Table~\ref{SU3table}.\footnote{Note again that while we use some of the same letters as in Sections~\ref{Sec:SO3Janus} and \ref{Sec:SU2U1} to denote these scalar fields, they correspond to different scalars in the maximal gauged supergravity.}
\begin{table}[t]
	\centering
	\begin{tabular}{@{\extracolsep{25 pt}}cc}
	\toprule
	\noalign{\smallskip}
		Field & $\SU(4)\times\U(1)_S$ rep \\
	\noalign{\smallskip}
\midrule
\noalign{\smallskip}
		$\varphi$, $c$ & ${\bf 1}_4\oplus{\bf 1}_{-4}$ \\[4pt]
		$\chi$, $\omega$ & ${\bf 10}_{-2}\oplus \overline{\bf 10}_2$ \\
		\noalign{\smallskip}
		\bottomrule
	\end{tabular}
	\caption{\label{SU3table}The scalar truncation of the maximal supergravity in five dimensions relevant for holographic dual to ${\cal N}=1$ interfaces. }
\end{table}
The explicit parametrization of the coset is the same as in \cite{Bobev:2019jbi}, but now embedded in the maximal theory as described in Appendix \ref{App: N=1 BPS eqs}.
%
%
For the Janus solutions of interest here we can consistently truncate out the fermions, gauge fields, and two-forms in the supergravity theory. The resulting Lagrangian is the same as in \eqref{Eq:5Dlagrangian},
where the scalar kinetic terms are  determined by  the matrix $M=U^TU$ and are presented \eqref{N1kinterms} in Appendix~\ref{App: N=1 BPS eqs}.  As before,  the scalar 27-bein, $U$, given in \eqref{Uforsu2}, has the same factorized structure as in \eqref{Umso3} and \eqref{Umsu2}.
The potential and superpotential are
\begin{equation}\label{Eq: (super)potential N=1}
\mathcal P = \frac{1}{2}  (\partial_\chi W)^2 - \frac{4}{3}\,W^2~,\qquad W =-\f{3g}{2}\cosh^2\chi~.
\end{equation}
The derivation of the BPS equations is outlined in Appendix \ref{App: N=1 BPS eqs}, see also \cite{Clark:2005te,Suh:2011xc}. The spin-1/2 variations lead to three BPS equations: 
\begin{equation}\label{Eq: N=1 BPS1}
\begin{aligned}
(\chi')^2 =&\, \f14 (\partial_\chi W)^2 -\frac{\cosh^2\chi}{\cos^2\left( c+ 2\omega \right)}(\varphi')^2~,\\
\omega' =&\, \sinh^2\varphi\,c'~,\\
\sinh2\varphi\,c' =&\, -2\tan(c+2\omega)\varphi',
\end{aligned}
\end{equation}
while the spin-3/2 variations yield the additional two:
\begin{equation}\label{Eq: N=1 BPS2}
\begin{aligned}
A' =\, -\frac13 \coth\chi\, (\chi')~,\qquad  \varphi' =\, 3\e^{-A}\cos(c+2\omega)\text{sech}\chi\tanh\chi~.
\end{aligned}
\end{equation}

Note that the structure of these equations differs somewhat from what we have encountered is Sections~\ref{Sec:SO3Janus} and \ref{Sec:SU2U1}. First, the differential equation for the dilaton, $\varphi$, does not come from the spin-1/2 variation. Secondly, there is no algebraic equation for the metric function, $A$. 
However, the analysis of these BPS equations proceeds in a similar fashion as in previous sections. 

The last two equations in \eqref{Eq: N=1 BPS1} produce the familiar integral of motion
\begin{equation}\label{eq:JN1Janus}
{\cal J} = \sinh2\varphi\,\sin(c +2 \omega)\,.
\end{equation}
The other integral of motion, $\mathcal I $, takes the form
\begin{equation}
\mathcal I =\frac{9 g^2}{5^{5/3}} \rme^{2A} \sinh^{2/3} \chi ~.
\end{equation}
The solution can be reduced to a classical mechanics problem, as in \eqref{Eq: N=4 classmech} and \eqref{eq:VeffN2},
\begin{equation}
\f{4}{g^2}(X')^2 + V_\text{eff} = 0~,
\end{equation}
where the convenient choice of $X$ is
\begin{equation}\label{}
X\eql -{1\over 3}\log \sinh \chi\,.
\end{equation}
The effective potential is then give by
\begin{equation}\label{Eq: N=1 Effective potential}
V_{\text{eff}} = 4\, \rme^{-2 X} \left( \frac{9}{5^{5/3}\mathcal I} - \rme^{-4 X} \cosh^2 3 X \right)~.
\end{equation}
The AdS$_5$ vacuum is found at
\begin{equation}
X\rightarrow +\infty~, \quad \text{such that} \quad \chi \rightarrow 0, \quad A\rightarrow +\infty ~.
\end{equation}

Again there is a static solution for $\mathcal I=1$ with $6X=\log 5$, which is discussed in Section~\ref{subsec:N1Jfold}. The non-static Janus solutions are only regular when $0 < \mathcal I \le 1$. In this range the classical mechanics problem is solved by \eqref{classSol}. The solutions represent a particle coming in from infinity, bouncing off the potential at the turning point $r_\text{tp}=r(X_{\text{tp}})$, where $V_{\text{eff}}(X_{\text{tp}})=0$, and returning back to infinity. We choose coordinates such that $r_\text{rp}=0$. For these solutions the remaining system of BPS equations collapses to a single separable differential equation for $\varphi(X)$, 
\begin{equation}
\frac{\sinh 2\varphi}{\sqrt{\sinh^2 2\varphi - \mathcal J^2}} \frac{\dd \varphi}{\dd X} =\pm \frac{9 \rme^{-X}}{5^{5/6}\sqrt{\mathcal I }\cosh 3X }\frac{1}{\sqrt{-V_{\text{eff}}}}\,.
\end{equation}
This equation does not admit an analytic solution in terms of elementary functions, but it can be integrated in quadratures and analyzed numerically. The five-dimensional dilaton takes the same form as in \eqref{Eq: N=4 dilatonsol}  and \eqref{eq:coshvarphiN2},
\begin{equation}
\cosh2\varphi =\cosh2F+\f12\e^{-2F}{\cal J}^2~, \qquad 
\end{equation}
with the function $F$ given by
\begin{equation}
 F = F_0 \pm \int\limits_{X_{\text{tp}}}^X \frac{9 \rme^{-x}}{5^{5/6}\sqrt{\mathcal I }\cosh 3x } \frac{\dd x}{\sqrt{-V_{\text{eff}}(x)}}~.
\end{equation}
This integral has to be performed numerically. 

\begin{figure}
\centering
\begin{subfigure}{.5\textwidth}
  \centering
  \includegraphics[width=0.95\textwidth]{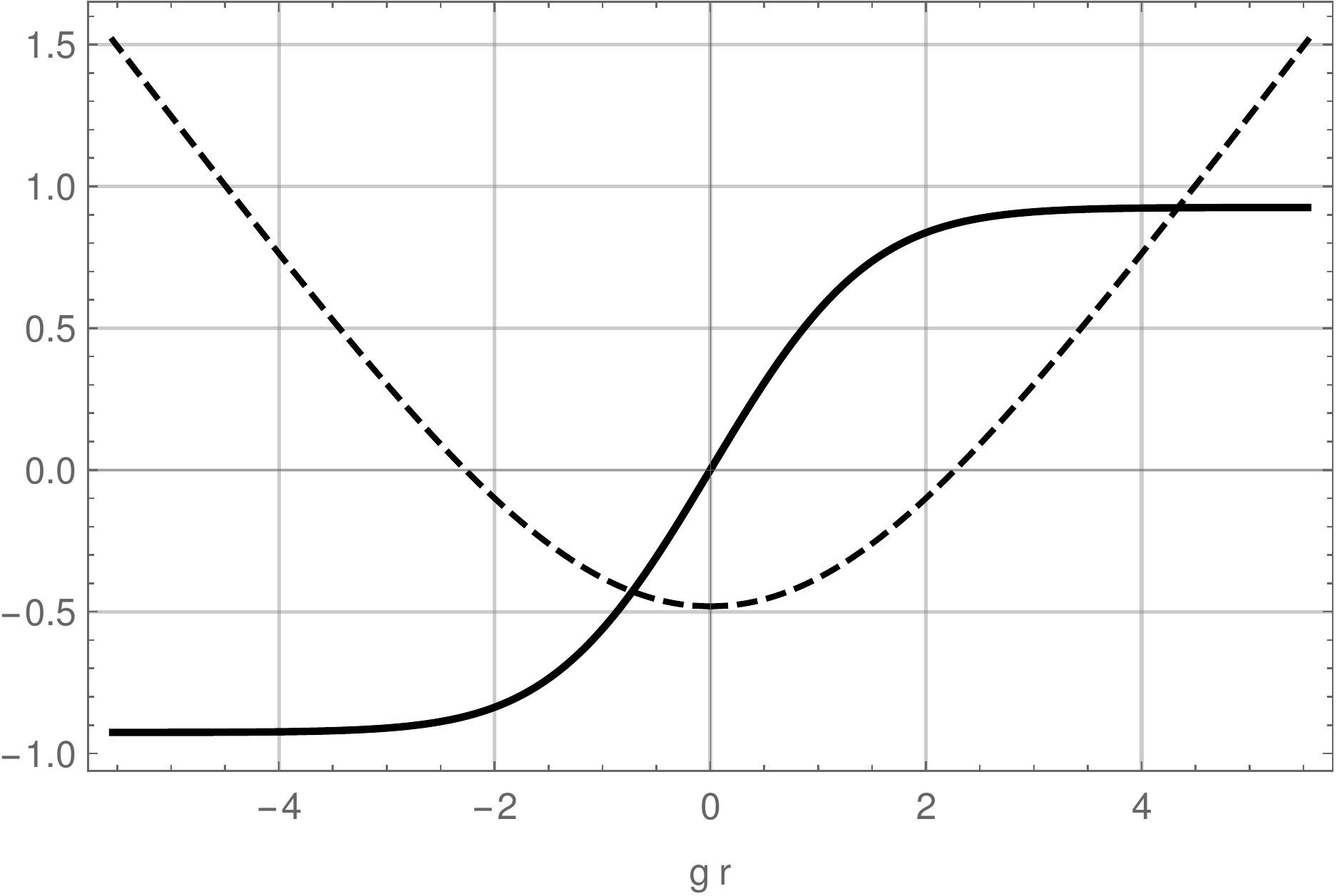}
\end{subfigure}%
\begin{subfigure}{.5\textwidth}
  \centering
  \includegraphics[width=0.95\textwidth]{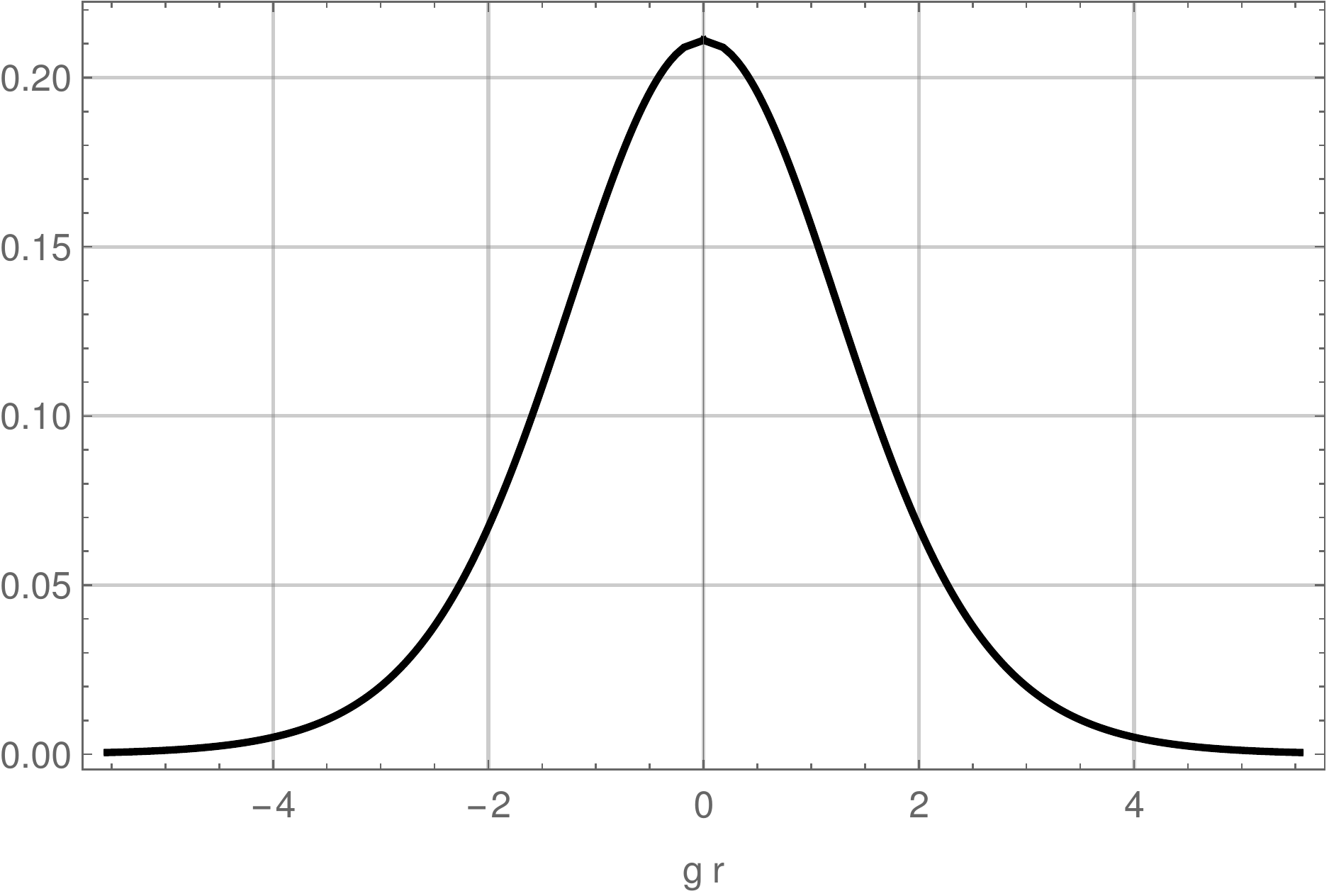}
\end{subfigure}
\caption{\label{SU3plot}A sample solution with ${\cal I}=4/5$. On the left pane we plot the functions $2(F(r)-F_0)$ (solid curve) that controls the solution for $\varphi$ and  $X-1$ (dashed curve) which is just a constant shift of the warp factor $A$. On the right pane we plot $\sinh\chi$.}
\end{figure}

Once more the axion is given by, see \eqref{Eq: N=4 csol} and \eqref{axionN2},
\begin{equation}\label{}
\cos^2(c - c_0) = \f{\sinh^22\varphi-{\cal J}^2}{(1+{\cal J}^2)\sinh^22\varphi}~.
\end{equation}
The final step is to determine the scalar $\omega$ which can be done using \eqref{eq:JN1Janus}. In Figure~\ref{SU3plot} we display a sample numerical Janus solution. 

The interpretation of the five constants which determine this family of $\N=1$ Janus solutions is the same as the one discussed below \eqref{eq:N4Janusconst} for the $\mathcal{N}=4$ interface. The only difference is that there is no operator of dimension $2$ sourced in the dual $\mathcal{N}=4$ SYM theory and thus the integration constant $\mathcal{I}$ controls only the change in the asymptotic value of the gauge coupling as well as the source for the dimension $3$ operator dual to the scalar field $\chi$.
%

\subsection{The ten-dimensional Janus}

This five-dimensional Janus solution can be uplifted to type IIB supergravity. The ten-dimensional metric is
\begin{equation}\label{eq:10dmetgen}
\dd s^2_{10} = \cosh\chi\,\dd s^2_5 + \frac{4}{g^2}\Big(\f{\dd s^2_{\mathbb{CP}^2}}{ \cosh\chi}+\cosh\chi~ \zeta^2\Big)\,,
\end{equation}
where $\dd s_5$ is the five-dimensional metric discussed in the previous section. The form $\zeta= \dd \phi + \sigma$ forms a $\U(1)$ bundle over the $\mathbb{CP}^2$ base, for which the K\"ahler form is given by $J=\dd \sigma$. Additionally, since the base is K\"ahler-Einstein one can construct a holomorphic $(2,0)$-form, $\Omega$, such that
\begin{equation}
\Omega \w \bar \Omega = 2 J\w J~,\quad \text{and} \quad  \dd \Omega = 3 \rmi \sigma \w \Omega~.
\end{equation}
This  allows us to write the NS-NS and R-R forms in a compact manner. The two-forms are give by\footnote{We use slightly different conventions when parametrizing the scalar manifold when compared to \cite{Bobev:2019jbi} which means that the ten-dimensional solution is in a different $\SL(2,\mathbb{R})$ frame. The two solutions can be identified by sending $c$ here to $c-\pi/2$.}
\begin{equation}
C_2 - \tau B_2 = -\frac{4\rmi}{g^2}\frac{\rme^{-\rmi \omega} \tanh \chi}{\cosh \varphi - \rme^{\rmi c} \sinh \varphi}\rme^{3\rmi \phi} \Omega~.
\end{equation}
The four-form potential is
\begin{equation}\label{eq:C4N1Janus}
C_4 = \frac{16}{g^4}\dd \phi \w \sigma \w J~.
\end{equation}
Finally, the axio-dilaton can be written as
\begin{equation}
\tau = C_0 + \rmi \rme^{-\Phi} = \frac{\rmi - \sinh 2\varphi \sin c}{\cosh 2\varphi - \sinh 2\varphi \cos c}~.
\end{equation}
This ten-dimensional $\mathcal{N}=1$ Janus solution agrees with the one found in \cite{DHoker:2006vfr}. In particular we find that it is possible to set the type IIB axion, $C_0$, to vanish by performing a global $\SL(2,\mathbb{R})$ transformation of the solution. This is analogous to the discussion in Section~\ref{subsec:UCLAcomp}.

\subsection{An ${\cal N}=1$ J-fold}
\label{subsec:N1Jfold}

Just as in the $\N=4$ and $\N=2$ case we find that for $\mathcal I=1$ something special happens. Namely, the effective potential, see  \eqref{Eq: N=1 Effective potential}, has an extremum equal to zero, for which the scalar takes the value $6X=\log 5$, such that
\begin{equation}
\sinh \chi = \frac{1}{\sqrt{5}}~.
\end{equation}
We also set $\mathcal{J}=c_0 =0$ using the $\SL(2,\mathbb{R})_S$ symmetry of the five-dimensional theory.

The metric is given by
\begin{equation}
\dd s_5^2 = \frac{5}{9g^2}\left( 4\dd \rho^2 + 5\dd s_{\text{AdS}_4}^2 \right) ~,
\end{equation}
where we have applied the coordinate transformation $\rho =\frac{3 g }{2\sqrt{5}}r$. As for the other J-folds, we find that the dilaton is linear $\varphi = \rho + \varphi_0$. To make this background regular we periodically identify the $\rho$-coordinate. This periodic identification has to be accompanied by an $\SL(2,\mathbb Z)$ monodromy with the same matrix $\mathfrak J$ as in \eqref{Jnmatrix}. This procedure results in the AdS$_4$ J-fold background constructed in \cite{Bobev:2019jbi}. 

The free energy of the three-dimensional $
\mathcal{N}=1$ SCFT dual to this solution was computed in \cite{Bobev:2019jbi} and reads
\begin{equation}
\mathcal F_{S^3}=\sqrt{\frac{5^5}{3^6}}\frac{N^2}{4}\text{arccosh}(n/2)\,.
\end{equation}

For completeness we also present the ten-dimensional uplift of this $\mathcal{N}=1$ J-fold solution. The metric is
\begin{equation}
\dd s^2_{10} = \sqrt{\frac{5}{6}}\frac{2}{3g^2}\left(4\dd \rho^2 + 5\dd s_{\text{AdS}_4}^2  + 6\dd s^2_{4}+ \frac{36}{5}\zeta^2\right)\,.
\end{equation}
The axio-dilaton is
\begin{equation}
\tau = C_0 + \rmi \e^{-\Phi} = \frac{\cosh(2\varphi+\rho_0)+\rmi \sinh\rho_0}{\cosh2\varphi}\,.
\end{equation}
The two-form potentials can be written as
\begin{equation}
C_2 - \tau B_2 = -\frac{2}{g^2}\frac{\sqrt{\tfrac{2}{3}\sinh\rho_0}}{\cosh\varphi+\rmi \sinh\varphi}e^{ 3\rmi\phi}\Omega~.
\end{equation}
The R-R four-form potential is the same as in the Janus solution \eqref{eq:C4N1Janus}.

\section{Conclusions}
\label{sec:conclusion}

In this paper, we constructed supergravity solutions in five and ten dimensions, which are holographically dual to three classes of Janus interfaces in $\mathcal{N}=4$ SYM  studied in \cite{DHoker:2006qeo}. These interfaces preserve three-dimensional $\mathcal{N}=4$, $\mathcal{N}=2$, and $\mathcal{N}=1$ supersymmetry, respectively. We  also found that with each of these Janus solutions one can associate an AdS$_4$ J-fold background of type IIB string theory, which is dual to a three-dimensional SCFT. Our results lead to some open questions and suggest several directions for future work.

We have focused on constructing $\mathcal{N}=2$ and $\mathcal{N}=1$ Janus solutions that in 
the classification of \cite{DHoker:2006qeo} are dual to the interfaces  with the maximal allowed flavor group. Using similar methods as in this paper it should be possible to construct the Janus solutions dual to the $\mathcal{N}=2$ interfaces with $\U(1)\times\U(1)$ global symmetry \cite{DHoker:2006qeo}, as well as $\mathcal{N}=1$ Janus solutions with no flavor symmetry. It is natural to expect that to each of these Janus solutions one can associate a corresponding J-fold background. It will certainly be interesting to find all these supergravity solutions explicitly. Note that in the classification of supersymmetric Janus interfaces in \cite{DHoker:2006qeo}, the $\theta$-term in the SYM Lagrangian was omitted. Given the importance of S-duality in  \cite{Gaiotto:2008sa,Gaiotto:2008sd,Gaiotto:2008ak} as well as the J-fold construction discussed in this paper, it might be interesting to revisit the analysis of \cite{DHoker:2006qeo} and look for novel superconformal interfaces in the presence of the $\theta$-term.

All our Janus solutions are smooth backgrounds of type IIB supergravity. As first pointed out in \cite{DHoker:2007zhm,DHoker:2007hhe}, one can find more general $\mathcal{N}=4$ AdS$_4$ solutions of the same type by allowing for explicit D-brane sources. These solutions are dual to superconformal interfaces in $\mathcal{N}=4$ SYM with extra degrees of freedom localized on the defect which correspond to the open-string excitations associated with the additional branes. Constructing similar Janus solutions with explicit brane sources and $\mathcal{N}=2$ and $\mathcal{N}=1$ supersymmetry is very difficult due to the  complexity of the type IIB supergravity BPS equations. The explicit Janus backgrounds we found in Section~\ref{Sec:SU2U1} and Section~\ref{Sec:SU3} may provide a useful starting point for generating such supergravity solutions by eschewing a full classification attempt and employing a suitable Ansatz for the supergravity fields.

The $\mathcal{N}=2$ and $\mathcal{N}=1$ J-fold AdS$_4$ solutions suggest the existence of a large class of new three-dimensional SCFTs. In Section~\ref{subsec:N2SCFT} we offered some suggestions as to what   the SCFT dual to the $\mathcal{N}=2$ J-fold configuration might be. It would be most interesting to establish this holographic correspondence more rigorously and also to extend it by studying more general J-fold configurations with explicit 5-brane sources as in \cite{Assel:2018vtq}. To this end one may utilize the results of \cite{Hashimoto:2014vpa,Hashimoto:2014nwa} on $\tfrac{1}{4}$-BPS boundary conditions in $\mathcal{N}=4$ SYM. The SCFTs with $\mathcal{N}=1$ supersymmetry are even more mysterious due to the low amount of supersymmetry, which prevents the use of many exact methods for computing physical observables.  Generalizations to include 5-brane sources in the string theory setup will require the study of $\tfrac{1}{8}$-BPS boundary conditions in $\mathcal{N}=4$ SYM, which is also a challenging task.

We have clearly shown the utility of lower-dimensional gauged supergravity theories arising as consistent truncations from string and M-theory to construct holographic duals to interfaces in supersymmetric QFTs. This approach can be effectively generalized to the study of defects and interfaces in the ABJM theory. Some explicit examples of supersymmetric Janus interface solutions in four-dimensional $\SO(8)$ gauged supergravity were found in \cite{Bobev:2013yra}. Before embarking on constructing further examples of similar Janus solutions it is desirable to establish a field theory classification, analogous to the one in \cite{DHoker:2006qeo}, for all superconformal interfaces in ABJM. Finally we would like to point out that gauged supergravity has proven to be a useful tool in the study of supersymmetric spatially modulated phases in the ABJM theory. Several examples of these ``susy Q'' solutions were constructed in four-dimensional gauged supergravity \cite{Kim:2018qle,Gauntlett:2018vhk,Arav:2018njv} and it will be very interesting to study their analogues in five-dimensions using the consistent truncations discussed in this paper.

\section*{Acknowledgements}
We are grateful to Mat Bullimore, Stefano Cremonesi, Adolfo Guarino, Noppadol Mekareeya, Wolfger Peelaers, and Silviu Pufu for useful discussions. NB and KP would like to thank Nick Warner for initial collaboration on this project and many helpful discussions. NB is supported in part by an Odysseus grant G0F9516N from the FWO. FFG is a Postdoctoral Fellow of the Research Foundation - Flanders (FWO). KP is supported in part by DOE grant DE-SC0011687. MS is supported by the National Research Foundation of Korea under the grant NRF-2019R1I1A1A01060811. The work of JvM is supported by a doctoral fellowship from the Research Foundation - Flanders (FWO). NB, FFG and JvM are also supported by the KU Leuven C1 grant ZKD1118 C16/16/005. KP would like to thank the ITF at KU Leuven for hospitality during two visits in the course of this project.

\appendix

\section{Derivation of the BPS equations}
\label{appendixSO3}
In this appendix we describe how the three consistent truncations of the $\SO(6)$ gauged maximal supergravity and the associated BPS equations are obtained. We follow the conventions of  \cite{Gunaydin:1985cu} except that we work in mostly plus signature.

In order to specify the generators of interest for our truncations within $\mathfrak{e}_{6(6)}$, we use an explicit representation of $\mathfrak{e}_{6(6)}$ in the so-called $\SL(6,\RR)\times \SL(2,\RR)$ basis given by the $27\times 27$ real matrices in   equation (A.36) in \cite{Gunaydin:1985cu}
\begin{equation}\label{GRWXdef}
\mathfrak{X}(\Lambda^I_{\phantom{I}J}, \Lambda^\alpha_{\phantom{\alpha}\beta},\Sigma_{IJP\beta}) = \begin{bmatrix}
-4\Lambda^{[M}_{\phantom{[M}[I}\delta^{N]}_{\phantom{N]}J]}& \sqrt{2}\Sigma_{IJP\beta}\\[6 pt]
\sqrt{2}\Sigma^{MNK\alpha} & \Lambda^K_{\phantom{K}P}\delta^{\alpha}_{\phantom{\alpha}\beta}+ \Lambda^\alpha_{\phantom{\alpha}\beta}\delta^{K}_{\phantom{K}P}
\end{bmatrix}~,
\end{equation}
where capital Latin indices run from 1 to 6 and lower case Greek indices run from 1 to 2. Here $\Lambda^I_{\phantom{I}J}$ and $\Lambda^\alpha_{\phantom{\alpha}\beta}$ are $\sl(6,\mathbb{R})$ and $\sl(2,\mathbb{R})$ generators, respectively, and $\Sigma_{IJP\beta}$ is real and completely antisymmetric in $IJK$. Indices on the $\Sigma$-tensor are raised using the $\SL(6,\mathbb{R})$ and $\SL(2,\mathbb{R})$ invariant tensors:
\begin{equation}
\Sigma^{IJK\alpha} = \f{1}{6}\varepsilon^{\alpha\beta}\varepsilon^{IJKLMN}\Sigma_{LMN\beta}~.
\end{equation}
The compact generators are those for which $\Lambda^I_{\phantom{I}J}$ and $\Lambda^\alpha_{\phantom{\alpha}\beta}$ are antisymmetric and $\Sigma^{IJK\alpha}$ is anti-selfdual; $\Sigma^{IJK\alpha}=-\Sigma_{IJK\alpha}$. For  the non-compact ones, $\Lambda^I{}_J$ and $\Lambda^\alpha{}_\beta$  are symmetric and traceless and $\Sigma^{IJK\alpha}=\Sigma^{IJK\alpha}$ is selfdual.

All three truncations discussed in this paper make use of the three generators of $\sl(2,\mathbb{R})$ spanned by $\Lambda^\alpha_{\phantom{\alpha}\beta}$. For convenience we denote these generators by
\begin{equation}\label{trsdef}
\mathfrak{t}=\mathfrak{X}(0,\sigma_1,0)~,\quad \mathfrak{r} = \mathfrak{X}(0,\rmi\sigma_2,0)~,\quad  \mathfrak{s}=\mathfrak{X}(0,\sigma_3,0)~,
\end{equation}
where we used the notation introduced in \eqref{GRWXdef} and $\sigma_i$ are the Pauli matrices. The $\SL(2,\mathbb{R})/\SO(2)$ scalar coset spanned by the axion and dilaton appears in all of our truncations and in all cases we parametrize that submanifold in the same way. To simplify our subsequent discussion we define the matrix
\begin{equation}
U_\text{dilaton} = \e^{-c \,\mathfrak{r}/2}\cdot \e^{\varphi \,\mathfrak{t}}\cdot \e^{(c/2+\pi/4)\mathfrak{r}}~,
\end{equation}
which will be utilized when we parametrize the scalar manifolds of the three truncations. The appearance of $\pi/4$ in the last exponent can be removed by a global $\SL(2,\mathbb{R})_S$ transformation, however we include it to simplify the form of the ten-dimensional uplifted solutions.

\subsection{The $\N=4$ Janus}
\label{app:N4}

In this appendix we present the derivation of the BPS equations in Section \ref{ssec:so3janus}. We write the generators of E$_{6(6)}$ that commute with $\SO(3)\times\SO(3)$ embedded in $\SO(6)$ as in \eqref{Eq: N4symmetry}. First we have the  $\sl(2,\mathbb{R})$-triple $(\mathfrak{t},\mathfrak{r},\mathfrak{s})$ as defined in \eqref{trsdef}. Second we have a ${\bf 20}'$ generator inside $\sl(6,\mathbb{R})$:
\be
\mathfrak{g}_\alpha = \mathfrak{X}(\Lambda^I_{\phantom{I}J}, 0,0)\quad\text{where}\quad \Lambda^I_{\phantom{I}J} = 
\left(\begin{array}{cc}
{\bf 1}_{3\times 3} & 0 \\ 
0 & - \,{\bf 1}_{3\times 3}
\end{array}\right)~.
\ee
Finally we have four generators with non-zero $\Sigma_{IJK\alpha}$. Two of those are non-compact, denoted by $\mathfrak{g}_{\chi_1}$ and $\mathfrak{g}_{\chi_2}$, and two are compact, denoted by $\mathfrak{r}_{\chi_1}$ and $\mathfrak{r}_{\chi_2}$. Explicitly, these are constructed from 
\begin{equation}
\begin{split}
\mathfrak{g}_{\chi_1},~\mathfrak{r}_{\chi_1}&:\quad\Sigma_{1231}=-\Sigma_{4561}=\sqrt{2}~,\\
\mathfrak{g}_{\chi_2},~\mathfrak{r}_{\chi_2}&:\quad\Sigma_{1231}=\Sigma_{4561}=\sqrt{2}~,
\end{split}
\end{equation}
with the same values of other components of the $\Sigma$-tensor related by symmetry and duality, before inserting them into \eqref{GRWXdef}. It is easy to verify that these eight generators span $\sl(3,\mathbb{R})$, and the three compact generators; $(\mathfrak{r},\mathfrak{r}_{\chi_1},\mathfrak{r}_{\chi_2})$ generate the compact $\SO(3)$ group that appears in the denominator of \eqref{Eq: N=4 Scalar manifold}. 

We parametrize the scalar coset in terms of these generators as follows
\begin{equation}\label{N4cosetparam}
U = \e^{\chi \,\mathfrak{g}_{\chi_1}}\cdot \e^{-\omega\, \mathfrak{r}}\cdot \e^{\alpha\,\mathfrak{g}_\alpha}\cdot U_\text{dilaton}~.
\end{equation}
Once $U$ has been specified it is a simple task to compute the matrix $M=U^T U$ and the kinetic terms using \eqref{Eq:5Dlagrangian}:
\begin{equation}\label{N4kinterms}
\begin{aligned}
  {1\over 24} & \,{\rm Tr} \, \partial_\mu M\partial^\mu M^{-1} =\\&- \f34 (5+3\cosh8\chi)(\alpha')^2 - 8(\chi')^2 - 2\sinh^22\chi(\omega')^2 + 4\sinh^2\varphi\,\sinh^22\chi\,(c')(\omega')\\
&-\f18\big(11+4\cosh4\chi+\cosh8\chi+8\cos(2c+4\omega)\sinh^42\chi\big)\Big[(\varphi')^2+\f14\sinh^22\varphi\,(c')^2\Big]\\
&-2\sinh^2\varphi\,\sinh^22\chi\,(\sinh^2\varphi-\cosh^2\varphi\,\sinh^22\chi\,\cos(2c+4\omega))(c')^2\\
&+2\sinh2\varphi\,\sinh^22\chi\,\sin(c+2\omega)\Big( 3\cosh^22\chi\,(\alpha') + \sinh^22\chi\cos(c+2\omega)(\varphi') \Big)(c')\\
&-3\cos(c+2\omega)\sinh^24\chi\,(\varphi')(\alpha')~.
\end{aligned}
\end{equation}
Our scalar parametrization was carefully chosen such that none of the phase angles, $c$ or $\omega$ appear in the $W$-tensors of \cite{Gunaydin:1985cu}. In particular from the $W_{ab}$ tensor we can extract the superpotential \eqref{Eq:SU2U1superpotential} that satisfies \eqref{SO3SO3pot} where the potential is just 
\be
{\cal P} = -\f{3g^2}{4}\left(3+\cosh4\alpha\,\cosh4\chi\right)~.
\ee

We are interested in supersymmetric solutions of the equations of motion. This means that we look for Killing spinors $\epsilon_a$ for which the supersymmetry variations\cite{Gunaydin:1985cu}
\begin{equation}\label{N8susyvars}
\begin{aligned}
\delta \chi_{abc} &=\, -\rmi\sqrt{2}\left(\gamma^\mu P_{\mu abcd}\epsilon^d + \f{3g\rmi}{2}W_{d[abc]|}\epsilon^d\right)~,\\
\delta \psi_{\mu a}&=\, \nabla_\mu\epsilon_a + Q_{\mu\,a}^{\p{\mu\,a}b}\epsilon_b - \f{g\rmi}{6} W_{ab}\gamma_\mu\epsilon^b~,
\end{aligned}
\end{equation}
vanish. Here $W_{ab}=W^c_{\p{c}acb}$, indices are raised an lowered with the symplectic matrix $\Omega_{ab}$. We refer to \cite{Gunaydin:1985cu} for further details. The procedure for finding the BPS equations starts by finding the eigenvectors of $W_{ab}$ that correspond to the superpotential and its complex conjugate 
\begin{equation}
g W_{ab} \eta^b_{(s)} = { W}\eta^a_{(s)}\,,\quad g W_{ab} \eta^b_{(\bar s)} = \overline{ W}\eta^a_{(\bar s)}\,,
\end{equation}
with $s,\bar s=1,2,3,4$. Then we can write $\epsilon^a = \eta^a_{(s)}\varepsilon^s + \eta^a_{(\bar s)}\varepsilon^{\bar s}$ and the vanishing of the supersymmetry variations reduce to equations for $\varepsilon^{s,\bar s}$. The spin-1/2 equation reduces to a condition on the spinor
\begin{equation}
\rmi\big(\alpha'-\sec(c+2\omega)\varphi'\big)\gamma_r\epsilon^s = \f{1+\sqrt{2}}{6}\partial_\alpha\overline{ W}\epsilon^{\bar s}~.
\end{equation}
This projector can be used to derive BPS equations for the scalars which appear in \eqref{Eq: N=4 BPS}. The spin-3/2 leads to a conformal Killing spinor equation on AdS$_4$
\begin{equation}\label{SO3dereps}
\nabla_i \epsilon^s = \f{\rmi(1+\sqrt{2})}6\overline{ W}\gamma_i\epsilon^{\bar s}~,\qquad \nabla_i \epsilon^{\bar s} = -\f{\rmi(-1+\sqrt{2})}6{ W}\gamma_i\epsilon^{s}~,
\end{equation}
where $i$ is an index along the AdS$_4$ slice. Finally we have a differential equation for $\epsilon^{s,\bar s}$ along the coordinate $r$ which fixes the radial dependence of the spinor. We simplify \eqref{SO3dereps} by writing the covariant derivative in terms of the derivative on ``unwarped'' AdS$_4$:
\begin{equation}
\nabla_i \epsilon^s =\tilde\nabla_{\hat \imath} \epsilon^s -\f12 A' \gamma_r\gamma_{i}\epsilon^s~,
\end{equation}
where ${\hat \imath}$ is an index on the unit-radius AdS$_4$.
A conformal Killing spinor $\eta$ on AdS$_4$ satisfies
\begin{equation}
\tilde\nabla_{\hat\imath}\eta = \kappa\f{\rmi}{2}  \gamma_5\gamma_{\hat\imath}\eta~,
\end{equation}
where $\gamma_5 = \gamma_{\hat0\hat1\hat2\hat3}=\gamma_r$ is the chirality operator in five dimensions and $\kappa^2 = 1$ is an arbitrary sign. Using this in \eqref{SO3dereps} we obtain
\begin{equation}
\left(A'-\kappa\f{\rmi}{\ell}\e^{-A} \right)\gamma_r\epsilon^s = \f{\rmi(\sqrt{2}+1)}3\overline{ W}\epsilon^{\bar s}~,\quad \left(A' -{\bar \kappa}\f{\rmi}{\ell}\e^{-A} \right)\gamma_r\epsilon^{\bar s} = -\f{\rmi(\sqrt{2}-1)}3{ W}\epsilon^{s}~.
\end{equation}
Consistency of the full system of equations requires $\kappa=-\bar \kappa$ and we are free to choose $\kappa = -1$. With this  we obtain the equations in \eqref{Eq: N=4 metricfunc}. 

\subsection{The $\N=2$ Janus}
\label{appendixsu2}

Here we outline the derivation of the BPS equations in Section \ref{Sec:SU2U1}. We write the generators of E$_{6(6)}$ that commute with $\SU(2)\times\U(1)$ embedded in $\SU(4)$ as in \eqref{SU2embedding}. First we have the  $\sl(2,\mathbb{R})$-triple $(\mathfrak{t},\mathfrak{r},\mathfrak{s})$ as defined in \eqref{trsdef}. Next we have a second $\sl(2,{\mathbb R})$-triple arising from the lower right corner of the $\sl(6,{\mathbb R})$ matrix:
\be
(\mathfrak{t}_{56},\mathfrak{r}_{56},\mathfrak{s}_{56}) = \mathfrak{X}(\Lambda^I_{\phantom{I}J}, 0,0)\quad\text{where}\quad \Lambda^I_{\phantom{I}J} = 
\left(\begin{array}{ccc}
{\bf 0}_{4\times 4} & 0  \\ 
0 & (\sigma_1,\rmi \sigma_2,\sigma_3)
\end{array}\right)~.
\end{equation}
One more ${\bf 20}'$ generator commutes with $\SU(2)\times\U(1)$ and is given by
\begin{equation}
\mathfrak{g}_{\alpha} = \mathfrak{X}(\Lambda^I_{\phantom{I}J}, 0,0)\quad\text{where}\quad\Lambda^I_{\phantom{I}J} = 
\left(\begin{array}{cc}
{\bf 1}_{4\times 4} & 0 \\ 
0 & -2 \,{\bf 1}_{2\times 2}
\end{array}\right)~.
\end{equation}
Finally we have four generators written in terms of the $\Sigma_{IJK\alpha}$. Two generators are non-compact and are denoted by $\mathfrak{g}_{\chi_{1,2}}$ and two are compact, denoted by $\mathfrak{r}_{\chi_{1,2}}$. These are specified by
\begin{equation}
\begin{split}
\mathfrak{g}_{\chi_1},~\mathfrak{r}_{\chi_1}&:\quad\Sigma_{1251}=\Sigma_{3451}=1~,\\
\mathfrak{g}_{\chi_2},~\mathfrak{r}_{\chi_2}&:\quad\Sigma_{1252}=\Sigma_{3452}=1~,
\end{split}
\end{equation}
with, as before, other components of the $\Sigma$-tensor  determined by  symmetry and duality. 
Altogether, these eleven generate $\SO(3,2)\times{\mathbb R}_+$ with a compact subgroup $\SO(3)\times\SO(2)$. 

We parametrize the scalar coset in terms of these generators as follows
\begin{equation}\label{appUsu2}
U = \e^{\alpha \,\mathfrak{g}_{\alpha}}\cdot\e^{\chi \,\mathfrak{g}_{\chi_1}}\cdot \e^{-\omega\, \mathfrak{r}}\cdot \e^{\alpha\,\mathfrak{g}_\alpha}\cdot  \e^{\lambda \,\mathfrak{t}_{56}}\cdot \e^{-\psi \,\mathfrak{r}_{56}/2}\cdot U_\text{dilaton}~.
\end{equation}
Then the kinetic terms  computed from \eqref{Eq:5Dlagrangian} are
\begin{equation}\label{N2kinterms}
\begin{aligned}
  {1\over 24} & \,{\rm Tr} \, \partial_\mu M\partial^\mu M^{-1} =\\
&-12(\alpha')^2 - 4(\chi')^2 - \sinh ^22 \chi\, (\omega')^2  - \f12(3+\cosh 4\chi)\big[(\lambda')^2+(\varphi')^2)\big]\\
&-  \sinh^2\varphi\,\left(\cosh2\varphi\, \cosh^22\chi+1\right)(c')^2 - \f14(\cosh4\lambda\,\cosh^22\chi-1)(\psi')^2\\
& -\sinh^22 \chi\,  \Big[-\sinh 2 \varphi \, \big(\sin (c+2 \omega)\lambda' - \f12 \sinh 2 \lambda \, \cos (c+2\omega ) \psi'\big)c'\\
&+2\cos (c+2 \omega ) \lambda' \varphi' +\sinh ^2\varphi\, (\cosh 2 \lambda\,\psi'  +2 \omega')c' + \sinh2\lambda\,\sin(c+2\omega)\varphi' \psi'\\
& + \cosh2 \lambda\,\omega' \psi' \Big]~.
\end{aligned}
\end{equation}

The parametrization of the coset in \eqref{appUsu2} is chosen such that none of the phase angles, $c$, $\omega$, or $\psi$ appear in the $W$-tensors of \cite{Gunaydin:1985cu}. In particular, from the $W_{ab}$ tensor we can extract the superpotential \eqref{Eq:SU2U1superpotential} that satisfies \eqref{SU2pot}, where the potential is
\be
{\cal P} = -\f{g^2}{4}\e^{-8\alpha}\left(1+4\e^{12\alpha}+8\e^{6\alpha}\cosh2\lambda~\cosh2\chi-\cosh4\lambda~\cosh^22\chi\right)~.
\ee

We now follow the same procedure as in Appendix~\ref{app:N4} and look for supersymmetric solutions of the equations of motion. The eigenvectors of $W_{ab}$ that correspond to the superpotential and its complex conjugate dictate which spinors we should consider 
\begin{equation}
g W_{ab} \eta^b_{(s)} = { W}\eta^a_{(s)}~,\quad g W_{ab} \eta^b_{(\bar s)} = \overline{ W}\eta^a_{(\bar s)}\,,
\end{equation}
with $s,\bar s=1,2$. Then we can write $\epsilon^a = \eta^a_{(s)}\varepsilon^s + \eta^a_{(\bar s)}\varepsilon^{\bar s}$ and the vanishing of the supersymmetry variations reduce to equations for $\varepsilon^{s,\bar s}$. The spin-1/2 equations reduce to the following condition on the spinor
\begin{equation}
\rmi(\alpha')\gamma_r\epsilon^s = \f1{12}\partial_\alpha\overline{ W}\epsilon^{\bar s}~.
\end{equation}
This projector can be used to derive BPS equations for the scalars which appear in~\eqref{eq:abl}-\eqref{eq:ang}.

The spin-3/2 leads to a conformal Killing spinor equation on AdS$_4$
\begin{equation}\label{SU2dereps}
\nabla_i \epsilon^s = \f{\rmi}6\overline{ W}\gamma_i\epsilon^{\bar s}~,\qquad \nabla_i \epsilon^{\bar s} = -\f{\rmi}6{ W}\gamma_i\epsilon^{s}~,
\end{equation}
where $i$ is an index along the AdS$_4$ slice. Finally we have a differential equation for $\epsilon^{s,\bar s}$ along $r$ which fixes the radial dependence of the spinor. Using exactly the same steps as in the case of ${\cal N}=4$ Janus we obtain
\begin{equation}
\left(-\kappa\f{\rmi}{\ell}\e^{-A}  +  A' \right)\gamma_r\epsilon^s = \f{\rmi}3\overline{ W}\epsilon^{\bar s}~,\qquad \left(+\kappa\f{\rmi}{\ell}\e^{-A}  +  A' \right)\gamma_r\epsilon^{\bar s} = -\f{\rmi}3{ W}\epsilon^{s}~.
\end{equation}
We have introduced the sign $\kappa$ as before to label the conformal Killing spinor we use on AdS$_4$. We are once again free to choose $\kappa = -1$. With this  we obtain the equations in \eqref{metricfuncSU2}.

\subsection{The $\N=1$ Janus}
\label{App: N=1 BPS eqs}

Finally, we also outline the derivation of the BPS equations in Section~\ref{Sec:SU3}. We write the generators of E$_{6(6)}$ that commute with the $\SU(3)$ subgroup of $\SO(6)$. As before, we have the  $\sl(2,\mathbb{R})$-triple $(\mathfrak{t},\mathfrak{r},\mathfrak{s})$ defined in \eqref{trsdef}. Next we have the $\U(1)$ generator that commutes with $\SU(3)$ and,  in our parametrization, is given by
\begin{equation}
\mathfrak{g}_{\U(1)} = \mathfrak{X}(\Lambda^I_{\phantom{I}J}, 0,0)\quad\text{where}\quad\Lambda^I_{\phantom{I}J} = 
\rmi \left(\begin{array}{ccc}
\sigma_2 & 0 & 0\\ 
0 & \sigma_2 & 0\\
0 & 0 & \sigma_2
\end{array}\right)~.
\end{equation}
In addition, we have four $\mathbf{10}\oplus \mathbf{\overline{10}}$ generators. Two of these are non-compact and denoted by $\mathfrak{g}_{\chi_{1,2}}$ and two are compact, denoted by $\mathfrak{r}_{\chi_{1,2}}$. These are specified by
\begin{equation}
\begin{split}
\mathfrak{g}_{\chi_1},~\mathfrak{r}_{\chi_1}&:\quad\Sigma_{1351}=-\Sigma_{1461}=-\Sigma_{2361}-\Sigma_{2451}=1~,\\
\mathfrak{g}_{\chi_2},~\mathfrak{r}_{\chi_2}&:\quad\Sigma_{1352}=-\Sigma_{1462}=-\Sigma_{2362}=-\Sigma_{2452}=1~.
\end{split}
\end{equation}
In total, these eight generators span $\SU(2,1)$ with a compact subgroup $\SU(2)\times\U(1)$.

We parametrize the scalar coset in terms of these generators as follows
\begin{equation}\label{Uforsu2}
U = \e^{\chi \,\mathfrak{g}_{\chi_1}}\cdot \e^{-\omega\, \mathfrak{r}}\cdot U_\text{dilaton}~.
\end{equation}
The kinetic terms then takes the explicit form
\begin{equation}\label{N1kinterms}
\begin{split}
  {1\over 24}  \,{\rm Tr} \, \partial_\mu M\partial^\mu M^{-1} &  =
 -2(\partial \chi)^2 -\frac{1}{2} \sinh^22\chi (\partial\omega - \sinh^2\varphi~\partial c)^2\\ & \qquad 
-\frac{1}{2} \cosh^2\chi\big[4(\partial \varphi)^2 + \sinh^22\varphi(\partial c)^2\big]~.
\end{split}
\end{equation}
Just as before one can construct the spinors we are interested in by looking at the eigenvectors and eigenvalues of the $W_{ab}$ tensor of \cite{Gunaydin:1985cu}. This time around one finds that 
\begin{equation}
g W_{ab} \eta^b_{(s)}= W \eta^a_{(s)}~, \quad \text{with}\quad  s=1,2,~
\end{equation}
the eigenvalue is given by the superpotential in \eqref{Eq: (super)potential N=1}. The spinors can thus be decomposed as
\begin{equation}
\epsilon^a = \eta_{(s)}^a \varepsilon^s,
\end{equation}
such that the equations in \eqref{N8susyvars} have to be solved independently for $\varepsilon^1$ and $\varepsilon^2$. From the spin-$1/2$ variation in \eqref{N8susyvars} one can find the equations in \eqref{Eq: N=1 BPS1} and the projector
\begin{equation}\label{Eq: Projector1}
\gamma_r \varepsilon^1 = \Omega \varepsilon^2\, \qquad \gamma_r \varepsilon^2 = \bar \Omega \varepsilon^1\, ,
\end{equation}
where
\begin{equation}
\Omega = -\frac{2 \rmi}{3 g} \frac{1}{\sinh \chi}\left( \frac{\chi'}{\cosh \chi} - \rmi \frac{\varphi'}{\cos \left( c+ 2\omega \right)} \right)\,.
\end{equation}
The consistency of the projectors is ensured by the identity $\left| \Omega \right|^2 = 1$. The BPS equations for $\varphi'$ and $A'$ in equation \eqref{Eq: N=1 BPS2} are found from solving the spin-$3/2$ variations in \eqref{N8susyvars}, and imposing that the solution is compatible with the projector found in the spin-$1/2$ variations. 


\bibliography{JanusBIB}

\bibliographystyle{utphys}
	
\end{document}